\documentclass[11pt]{article}

\usepackage{amsmath}

\usepackage[ansinew]{inputenc}
\usepackage{color,graphicx}
\usepackage{wrapfig}
\usepackage{amssymb}
\usepackage{amsmath}
\usepackage{dsfont}
\usepackage{natbib}
\usepackage{breqn}
\usepackage{enumerate}
\usepackage{algorithm}
\usepackage[noend]{algpseudocode}
\usepackage{geometry}
\usepackage{tabulary}
\usepackage{caption}
\usepackage[english]{babel}
\usepackage[ansinew]{inputenc}
\usepackage{epsfig}
\usepackage{color,graphicx}
\usepackage{wrapfig}
\usepackage{amssymb,amsthm,amsmath}
\usepackage{dsfont}
\usepackage{natbib}
\usepackage{abstract}
\usepackage{enumerate}
\usepackage{comment}
\usepackage{algorithm}
\usepackage[noend]{algpseudocode}

\newtheorem{proposition}{Proposition}[]

\newcommand{\R}{\mathds{R}}
\newcommand{\N}{\mathds{N}}
\newcommand{\I}{\mathbb{I}}

\algrenewcommand\algorithmicrequire{\textbf{Input:}}
\algrenewcommand\algorithmicensure{\textbf{Output:}}

\newcommand{\ds}{\displaystyle}

\begin{document}
\sloppy


\title{Exact and computationally efficient Bayesian inference for generalized Markov modulated Poisson processes}

\author{F. B. Gon\c{c}alves$^{a,1}$ \and L. M. Dutra$^{b}$ \and R. W. C. Silva$^{a}$}

\date{}

\maketitle

\begin{center}
{$^a$ Universidade Federal de Minas Gerais\\
$^b$ Centro Federal de Educa\c{c}\~ao Tecnol\'ogica de Minas Gerais}
\end{center}

\footnotetext[1]{Address: Av. Antônio Carlos, 6627 - DEST, ICEx, UFMG -
31270-901, Belo Horizonte, Minas Gerais, Brazil. E-mail: fbgoncalves@est.ufmg.br}


\footnotetext[1]{Address: Av. Ant\^onio Carlos, 6627 - DEST, ICEx, UFMG -
31270-901, Belo Horizonte, Minas Gerais, Brazil. E-mail: fbgoncalves@est.ufmg.br}


\begin{abstract}
Statistical modeling of point patterns is an important and common problem in several areas. The Poisson process is the most common process used for this purpose, in particular, its generalization that considers the intensity function to be stochastic. This is called a Cox process and different choices to model the dynamics of the intensity gives rise to a wide range of models. We present a new class of unidimensional Cox process models in which the intensity function assumes parametric functional forms that switch among them according to a continuous-time Markov chain. A novel methodology is proposed to perform exact Bayesian inference based on MCMC algorithms. The term exact refers to the fact that no discrete time approximation is used and Monte Carlo error is the only source of inaccuracy. The reliability of the algorithms depends on a variety of specifications which are carefully addressed, resulting in a computationally efficient (in terms of computing time) algorithm and enabling its use with large data sets. Simulated and real examples are presented to illustrate the efficiency and applicability of the proposed methodology. A specific model to fit epidemic curves is proposed and used to analyze data from Dengue Fever in Brazil and COVID-19 in some countries.

\end{abstract}

\section{Introduction}

Point pattern statistical models aim at modeling the occurrence of a given event of interest in a given region, which is commonly interpreted as time in the unidimensional case. The most widely used point process model is the Poisson process (PP) in which the number of events in any region has Poisson distribution and is independent for disjoint regions.
The Poisson process dynamics is mainly determined by its intensity function (IF) and is called a homogeneous Poisson process when this function is constant. Cox processes are a statistically appealing generalization of the Poisson process that allow the intensity function to vary stochastically across the region under consideration. A variety of classes of Cox process models can be defined in terms of the stochastic dynamics that describes the variation of the intensity function. Several of those models have already been proposed in the literature, including non-parametric models in which the IF is described by a Gaussian process \citep*{moller,G&G} or a diffusion process \citep*{GLR}. A simple yet appealing class of models are the Markov modulated Poisson processes (MMPP) in which the IF follows a continuous time Markov chain (CTMC). This means that the IF is piece-wise constant with jumps having a Markovian dynamics. This class of models has been explored, under a statistical perspective, by different authors before. In particular \citet{fearnhead2006exact} and \citet{rao2013fast} propose exact (free of discretization error) Monte Carlo methodologies to perform Bayesian inference. Whilst the former scales with the number of observations, the latter scales with the number of changes in the IF and is, therefore, more computationally efficient.

This paper proposes a generalization of Markov modulated Poisson processes - called the generalized MMPP (GMMPP), that allows the intensity function to jump among different and pre-specified functional forms. The jumps are determined by a continuous time Markov chain but in a way that each state of the chain is associated to one functional form. The model is actually specified in a way that self-jumps of the IF are allowed, meaning that the IF may restart in the same functional form. Furthermore, each functional form is allowed to depend on unknown parameters and the starting value of the IF in one functional form may vary among different visits of the chain to that state. This construction offers a quite flexible yet parametric solution to model the IF of unidimensional Cox processes.

The proposed class of Cox processes is expected to fill a gap between Markov modulated Poisson processes and fully non-parametric Cox processes. Compared to the former, GMMPPs provide much more flexibility to model the IF dynamics. A MMPP model would typically require the use of CTMC with large state spaces with many state changes (short visits) which would seriously compromise model parsimony and, consequently, the implied computational cost. On the other hand, when compared to fully non-parametric approaches, for example, when the IF dynamics depends on a Gaussian process, GMMPPs are expected to provide a similar good fit in many cases but with huge gains in terms of computational cost.

An MCMC algorithm is developed to perform exact Bayesian inference for GMMPPs. It is exact in the sense that the devised Markov chain converges to the exact posterior distribution of all the unknown quantities of the model, including the IF. The algorithm builds upon the ideas introduced in \citet{rao2013fast} so that it scales with the number of Markov jumps and does not suffer massively when increasing the number of observations. Further non-trivial developments are proposed to circumvent the fact that a forward-filtering-backward-sampling (FFBS) cannot be used to sample one of the blocks of the Gibbs sampling as it is done in \citet{rao2013fast}. In fact, one of the ideas developed here can be used to further improve the algorithm of \citet{rao2013fast}. The proposed MCMC is computationally efficient in terms of computing time and, therefore, feasible to be used with very large data sets. This way, the main contributions of this paper are twofold: first, a novel class of parametric unidimensional Cox process which are flexible yet parsimonious is propose and, second, a computationally efficient MCMC algorithm is proposed to perform exact Bayesian inference. The proposed methodology offers an appealing (and much cheaper) alternative to non-parametric Cox processes in a variety of problems in which the latter ought to be a suitable choice.

\begin{figure*}[!h]
\centering
\includegraphics[width=1\linewidth]{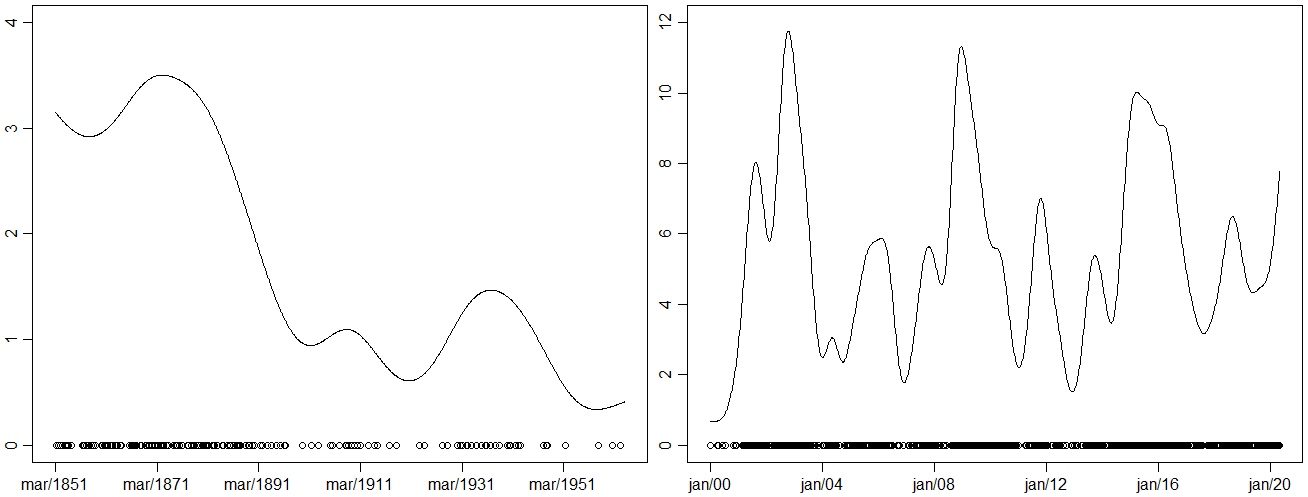}
\caption{Two real data sets and the respective empirical IF. Right: Coal mining disasters - time unit = year. Left: daily variation over $1\%$ in the BRLxUSD exchange rate - time unit is month.}\label{fig0}
\end{figure*}

In order to motivate the use of GMMPPs we consider two real data sets regarding coal mining disasters and exchange rate between Brazilian Real (BRL) and US Dollar (USD). For the former, each event represents an explosion that killed ten or more men in Britain. For the latter, each event represent a day in which the variation w.r.t. the previous day was greater than $1\%$. A kernel method \citep[see][]{diggle1d} is used to estimate the IF as it is shown in Figure \ref{fig0} and suggest that the IF ought to be well described, in both examples, by a GMMPP with an increasing and a decreasing line with varying starting values. Both examples are revisited in Section 5.

Finally, we propose a specific model to fit epidemic curves, allowing for asymmetry between the growth and decay behaviors. The model allows for simplifications in the proposed MCMC algorithm, which lead to reasonable computational times even for very large data sets - with more than 300 thousand observations.

This paper is organized as follows. Sections 2 and 3 of the paper present the GMMPP and the proposed MCMC algorithm, respectively. Section 4 explores some simulated examples to discuss the efficiency of the proposed methodology. In particular, the proposed methodology is compared to a non-parametric Cox process approach in terms of inference and computational cost. Finally, Section 5 applies the methodology to some real data sets. Two of them are the ones presented in Figure \ref{fig0} regarding coal mining disasters and the BRLxUSD exchange rate. A third example considers data sets regarding Dengue Fever epidemics in Brazil and the COVID-19 pandemic, for which a specific GMMPP model is carefully designed.

\section{Generalized Markov modulated Poisson processes}

Let $Y := \{Y(s)\}_{s \in \R^+}$ be a non-homogeneous Poisson process (NHPP) with intensity function $\lambda := \{\lambda(s)\}_{s \in \R^+}$ and consider $K$ functional forms $g_k$, $k=1,\ldots,K$, to be assumed by $\lambda$ along $\R^+$. These may be, for example, constants, increasing or decreasing lines, exponential functions, etc. The IF $\lambda$ switches among the different functional forms according to the transitions of a continuous-time Markov chain $X := \{X(s)\}_{s \in \R^+}$, with Q-matrix $Q_{\theta}$, initial distribution $\pi_0$ and state space $\{1:E\}:=\{1, 2, \cdots, E\}$, for $E\geq K$, where $\theta$ is the vector of parameters indexing $Q_{\theta}$. Furthermore, in its most general form, we allow the IF to switch from the same functional form to itself and have different starting values every time a functional form is revisited. We call the resulting process $Y$ a \emph{generalized Markov modulated Poisson process} with mathematical representation given as follows.

Define $T = (T_1, T_2, \cdots)$ as the jump times of $X$ and $Z = (Z_1,Z_2, \cdots)$ as the corresponding sequence of visited states, i.e. $Z_i = X(T_i)$, $i \in \N$, and $Z_0 = X(0)$. Then,
\begin{eqnarray}
&Y&\sim \mbox{NHPP}(\lambda), \label{eq_model_1} \\
&\lambda(s)& = g_{h_s}(s,\delta_s, R_s,\psi),\label{eq_model_2} \\
&X&\sim  \mbox{CTMC} (\pi_0, Q_{\theta}, E),  \label{eq_model_5} \\
&R_{[0,S)} &\sim  prior.  \label{eq_model_4}
\end{eqnarray}

For each $s$, the surjective function $h_s:=h(X(s)): \{1:E\} \rightarrow \{1:K\}$ assigns a functional form $g_k$ to each of the states of $X$ and $\delta_s:=\delta_s(X) = \max\limits_{i \geq 0}\{ T_i: T_i \leq s\}$. Additionally, $R_s := R_s(X)=\lambda(\delta_s)$ is the starting value of the IF at the last jump time up to $s$ and, finally, $\psi$ is a vector of parameters indexing the $g_k$'s. For example, suppose that $g_1$ is a straight line with inclination $\beta$, then, for a given $s$ such that the IF assumes the functional form $g_1$ at $s$, we have that $\lambda(s)=R_s+\beta(s-\delta_s)$. Naturally, the intensity function $\lambda$ is required to be non-negative. Formally, we deal with this issue by assigning zero to the density of $Y$ conditional on any trajectory of the IF that assumes negative values. The prior on the starting values $R_s$ is presented in Section \ref{sec_Binf}.

\begin{figure*}[!h]
\centering
\includegraphics[width=1\linewidth]{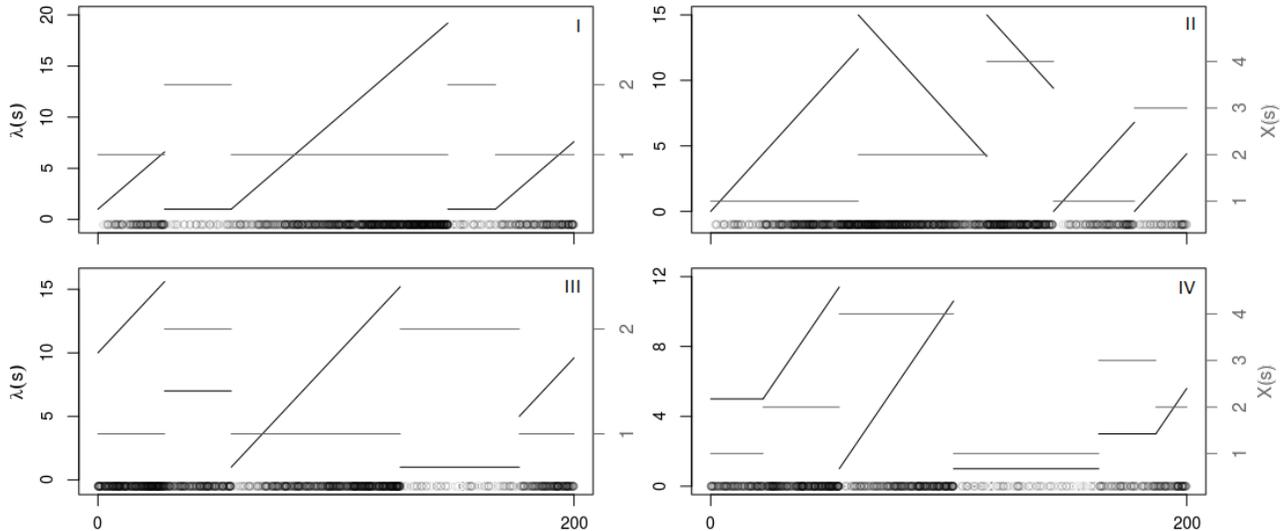}
\caption{Four examples of GMMPPs. I- two functional forms: constant and increasing line. No self-jumps allowed and no varying starting value. II- two functional forms: increasing and decreasing lines. Self-jumps allowed and no varying starting value. III- two functional forms: constant and increasing line. No self-jumps allowed and varying starting value. IV: two functional forms: constant and increasing line. Self-jumps allowed and varying starting value. Colour gray refers to the CTMC trajectory and the circles represent the events from the Poisson process.}\label{fig1}
\end{figure*}

In the simpler case in which the IF is not allowed to switch from each $g_k$ to itself, we set $h(k)=k$, $k=1,\ldots,K$. On the other hand if that feature is allowed, we set $h(k)=h(K+k)=k$, $k=1,\ldots,K$. In order to favor model identifiability in a statistical context, some entries of the $Q_{\theta}$ matrix are set to be zero so that a jump to the $k$-th functional form that is not a self-jump can only happen through the $k$-th state of $X$ and never through the $(K+k)$-th one. Figure \ref{fig1} illustrate the proposed class of models by presenting a realization of each of four different models.

\section{Bayesian inference}\label{sec_Binf}

We aim at performing inference for GMMPPs based on the observation of the process over a finite length time interval $[0,S]$. The proposed methodology is meant to be exact in the sense that no discrete time approximation of the original process is to be considered. In particular, we shall perform Bayesian inference via an MCMC algorithm that has the exact posterior distribution of all the unknown quantities in the model as its invariant distribution. As a result, Monte Carlo error is the only source of inaccuracy.

The main aim of the inference process is to obtain the posterior distribution of the intensity function and unknown parameters. Given the structure of the proposed class of models, this is equivalent to the distribution of $(Z_0,Z,T,\theta,\psi|y)$, where $y$ represents a realization of the process $Y$ in $[0,S]$. In order to fully specify each model under the Bayesian approach we need to assign a prior distribution to parameters $\theta$ and $\psi$. We define $\psi=(\psi_1,\ldots,\psi_K)$, where $\psi_k$ is the set of parameters indexing the $k$-th functional form, $\theta_{\cdot} = \{ \theta_k: k = 1, \cdots, K\}$ as the rates of the waiting times of $X$ and $\theta_{k \cdot} = \{ \theta_{kj}: j = 1, \cdots, K \mbox{ and } j \neq k \} $ as the transition probabilities from the states corresponding to the $k$-th functional form. The parameter vectors $\theta$ and $\psi$ are assumed to be independent \emph{a priori}. Independence among all the $\psi_k$'s and among the components of $\theta_{\cdot}$ and vectors $\theta_{k \cdot}$ is also assumed. In the case that no self-jumps are allowed, the full prior specification of $\theta$ is completed by setting, for $k=1,\dots,K$,
\begin{eqnarray}\label{eq_priors1}
\theta_k&\sim& Gamma(\alpha_k,\beta_k),\\
\theta_{k \cdot}&\sim& Dirichlet(\gamma_{k1},\ldots,\gamma_{kk-1},\gamma_{kk+1},\ldots,\gamma_{kK}).\nonumber
\end{eqnarray}

In the case in which self-jumps are allowed, each row of $Q_{\theta}$ has $K$ non-zero probabilities due to the restrictions imposed to have model identifiability. Moreover, the transition probabilities between the two states corresponding to the same functional form are the same and the transition probabilities between the $k_1$-th and $k_2$-th functional forms are the same whether moving from state $k_1$ to $k_2$ or from state $k_1+K$ to $k_2$. All this means that there are $K$ transition probabilities associated to each functional form and the vector of these probabilities are assumed to follow a Dirichlet distribution as follows. For the $k$-th functional form, $k=1,\dots,K$, we set
\begin{equation}\label{eq_priors2}
\theta_{k \cdot}\sim Dirichlet(\gamma_{k1},\ldots,\gamma_{kK}).
\end{equation}
Furthermore, the rate parameters of the waiting times are the same for the $k$-th and $(k+K)$-th states. In order to illustrate all the restrictions imposed to the $Q_{\theta}$ matrix, consider an example with three functional forms, all allowed to self-jump. The resulting $Q_{\theta}$ matrix is then given by

\begin{center}
$\left(
  \begin{array}{ccccccccccc}
    -\theta_1    & &\theta_1\theta_{12} & &\theta_1\theta_{13} & & \theta_1\theta_{11} & & 0          & &0           \\
    \theta_2\theta_{21} & &-\theta_2    & &\theta_2\theta_{23} & &0           & &\theta_2\theta_{22} & &0           \\
    \theta_3\theta_{31} & &\theta_3\theta_{32} & &-\theta_3    & &0           & & 0          & &\theta_3\theta_{33} \\
    \theta_1\theta_{11} & &\theta_1\theta_{12} & &\theta_1\theta_{13} & &-\theta_1   & &  0          & & 0           \\
    \theta_2\theta_{21} & &\theta_2\theta_{22} & &\theta_2\theta_{23} & & 0          & &-\theta_2    & & 0           \\
    \theta_3\theta_{31} & &\theta_3\theta_{32} & &\theta_3\theta_{33} & & 0          & & 0          & & -\theta_3    \\
  \end{array}
\right)$
\end{center}

Let $R$ be the set of starting values of the intensity function at all the jump times of $X$ in $[0,S]$. Note that the dimension of $R$ is random and depends on the number of jumps. Therefore, the prior distribution on $R$ is defined conditional on $(T,Z_0,Z)$, as follows:
\begin{eqnarray}
\pi(R|T,Z_0,Z)=\prod_{i=0}^{|T|}\pi(R_i|Z_i), \label{eq_prior_R_1} \\
\pi(R_i|Z_i=k)=\pi_k, \label{eq_prior_R_2}
\end{eqnarray}
where $|T|$ is the number of jumps in $[0,S]$ and $R_i$ is the starting value of the IF at $T_i$. The prior distribution in $(\ref{eq_prior_R_1})$-$(\ref{eq_prior_R_2})$ assumes a structure of conditional independence among all the starting values and identical distribution among all the starting values referring to the same functional form. Furthermore, in order to have a feasible MCMC algorithm, the constant functional form is the only one for which a continuous prior can be adopted, in particular, a Gamma distribution. For all the other forms, a discrete prior must be adopted and, unless useful information is available, we shall assume uniform discrete priors on supports chosen according to the scale of the IF. A more flexible approach is to set the size of the support of this discrete distribution and set the actual values as unknown and assuming a joint continuous prior.

\subsection{Model elicitation and identifiability}

The proposed GMMPP models offer a considerably flexible structure to model a variety of point process phenomena. This flexibility, however, gives rise to complex important issues that have great influence in the quality of the statistical analysis to be performed. More specifically, model and prior elicitation should be carefully performed to avoid identifiability problems and favor a reasonable model fit. Reliable prior information about the phenomenon under study should always be used for this purpose. Additional strategies may include preliminary analysis of the data and the use of informative priors. Regarding the former, one may obtain a non-parametric kernel estimate of the intensity function \citep[see][]{diggle1d} and use this to guide the choice for the functional forms and other features (self-jump, variable starting value) of the model. The use of informative priors should be considered based on the (prior) information acquired. For example, if few transitions are expected, the data would provide little information about the parameters indexing the $Q_{\theta}$ matrix. In this case, the information that few transitions are expected could be used to elicit informative priors for the mean waiting time parameters $\theta_i$ in terms of the scale of the model (magnitude of the waiting times). The same strategy may be used to set informative priors, also in terms of the scale, for the parameters $\psi$ indexing the adopted functional forms. Generally speaking, the proposed models do not aim at emulating non-parametric structures, which would imply the need for many functional forms with short visits to each one. This would compromise model identifiability and the computational cost. The actual aim of the proposed models is to provide good model fitting and prediction with high gains in terms of computational cost in situations in which a non-parametric structure for the IF is expendable.

All of the issues discussed above are explored in the simulated and real examples presented in Sections \ref{sec.sim} and \ref{sec.app}.

\subsection{Model augmentation and the MCMC algorithm}\label{ssec.mcmc}

The target posterior distribution of $(Z_0,Z,T,\theta,\psi|y)$ is highly complex, which suggests the use of MCMC as the most reasonable choice to perform inference. Developing an efficient algorithm, however, is not straightforward and ought to consider non-trivial techniques and algorithms to achieve that goal. We propose a model augmentation approach similar to the one proposed in \citet{rao2013fast} but with some adaptations to gain in computational efficiency. The model augmentation is based on the augmented representation of a CTMC proposed in \citet{hobolth2009simulation} and referred to as uniformization. The CTMC is represented as a discrete time Markov chain (DTMC) subordinated to a Poisson process. This means that the times of the DTMC, which has the same state space of the CTMC $X$, are defined by a Poisson process. The augmented component comes from the fact that the DTMC may have transitions between the same state. We shall refer to those type of transitions as virtual jumps.

The difference between our approach and the one in \citet{rao2013fast} is that we consider a non-homogeneous Poisson process instead of homogeneous one. The gain in efficiency due to the use of a non-homogeneous PP will be made clear further ahead in the text. The representation using non-homogeneous PP is also proposed in \citet{rao2012} in the context of inference for semi-Markov jump processes. Nevertheless, the authors do not provide a proof for the result.

Let us start by defining $K$ constants $\Omega_1,\ldots,\Omega_K$ such that $\Omega_k>|Q_k|$, where $Q_k$ is the $k$-th diagonal element of $Q_{\theta}$. Now let $V_0,V_1,\ldots$ be a sequence of discrete random variables on $\{1:E\}$ and $W=(W_1,W_2,\ldots)$ a sequence of random times on $\mathds{R}^+$. We define the following stochastic process:
\begin{align}
&V_0 \sim \pi_0, \label{eq_augCTMC_1} \\
&(W_1|V_0 = j) \sim \mbox{Exponential}(\Omega_{h(j)}), \label{eq_augCTMC_2} \\
&(V_{1}|V_{0} = j) \sim B_{j \cdot} = \textbf{1}_j + \frac{1}{\Omega_{h(j)}}Q_{j \cdot}, \label{eq_augCTMC_3}\\
&(W_{\ell}- W_{\ell-1}|V_{\ell-1} = j, W_{\ell-1})\label{eq_augCTMC_4} \\&\;\;\;\;\;\;\;\;\;\;\;\;\;\;\;\,\,\,\,\,\,\,\,\sim \mbox{Exponential}(\Omega_{h(j)}),  \nonumber\\
&(V_{\ell}|V_{\ell-1} = j) \sim B_{j \cdot} = \textbf{1}_j + \frac{1}{\Omega_{h(j)}}Q_{j \cdot}, \label{eq_augCTMC_5}
\end{align}
for $\ell=2,\dots,|W|$, where $Q_{j \cdot}$ is the $j$-th row of $Q_{\theta}$ and $B_{j \cdot}$ is a probability vector such that $\textbf{1}_j$ is a row vector of size $E$ with the $j$-th element being 1 and all the others being 0.

As it is stated in Proposition \ref{prop_1} below, the process $(V_0,V)$, where $V=(V_1,V_2,\ldots)$, subordinated to times $W$ is an alternative representation for the CTMC $X$. We shall refer to this process as the augmented CTMC. Note that the virtual times are an extra component that is not defined in the original definition of a CTMC. Finally, note that the result is valid for $\Omega_k\geq|Q_k|$ but, in order to use this representation in our MCMC context, we required the strict inequality. The equality implies in the almost surely non-existence of virtual jumps whilst these are crucial to establish the validity of the MCMC algorithm to be proposed, as it will be made clear further ahead in the text.

\begin{proposition}\label{prop_1}
For any $\Omega_k \geq |Q_k|$, the process $(V_0, V, W)$ defined in (\ref{eq_augCTMC_1})-(\ref{eq_augCTMC_5}) is a valid augmented representation of a continuous time Markov chain with initial distribution $\pi_0$ and Q-matrix $Q_{\theta}$.
\end{proposition}
\begin{proof}
See Appendix A.
\end{proof}

Consider now the augmented model that replaces the CTMC $X$ in the original model in (\ref{eq_model_1})-(\ref{eq_model_4}) by the augmented CTMC defined in (\ref{eq_augCTMC_1})-(\ref{eq_augCTMC_5}). We define $U$ and $T$ as the virtual and non-virtual jumps of the augmented CTMC, respectively.
The vector of all the unknown quantities in the augmented model is $\varphi=(W,U,T,V_0,V,\theta,\psi)$. This means that the aim of the inference procedure is to obtain the posterior distribution of $(\varphi|y)$. Note that there is a redundancy in the definition of $\varphi$ since $W=U\cup T$, nevertheless, that is required due to the particular sampling scheme to be adopted in the MCMC.

We design a Gibbs sampling to sample from the target posterior distribution. The blocking scheme and sampling algorithms to be adopted aim at simultaneously optimizing the convergence properties and computational cost of the Markov chain. We consider the following blocks: $(U,W,V_{W})$, $(V_0,V,U,T,R)$, $\theta$, $\psi$, where $V=V_W\cup V_T$ and $V_W$ and  $V_T$ are $V$ at times $W$ and $T$, respectively.

Before describing the algorithms to sample from each block, we present the joint density of $(Y,\varphi)$ which is useful to derive those algorithm since all the full conditional densities are proportional to this joint density. We have that
\begin{eqnarray}\label{eq_post_dens}
&&\pi(Y,\varphi) = \pi(Y|V_0, V , W, R, \psi)\\
&&\times\pi(V_0, V, T, U, W | \theta)\pi(R|V_0,V,T) \pi(\theta) \pi(\psi),\nonumber
\end{eqnarray}
where each density above is obtained w.r.t. some suitable dominating measure. The likelihood $\pi(y|V_0, V , W, R, \psi)$ is written w.r.t. the probability measure of a Poisson process with constant rate such that
\begin{eqnarray}\label{eq_likelihood}
\pi(y|V_0, V , W, R, \psi)&\stackrel{\varphi}{\propto}&\exp\left\{- \int_{0}^S \lambda(s) ds \right\}\nonumber\\
&&\times\prod_{n=1}^{N_S}\lambda(t_n),
\end{eqnarray}
where $N_S$ is the number of events from $y$ in $[0,S]$ and $t_n$ is the time of the $n$-th event. The densities $\pi(V_0, V, T, U, W | \theta)$, $\pi(R|V_0,V,T)$ and $\pi(\theta)$ can be obtained from (\ref{eq_augCTMC_1})-(\ref{eq_augCTMC_5}), (\ref{eq_prior_R_1})-(\ref{eq_prior_R_2}) and (\ref{eq_priors1})-(\ref{eq_priors2}), respectively. Finally, $\pi(\psi)$ is some suitable continuous density.

\subsubsection*{Sampling $(U,W,V_{W})$}

The block $(U,W,V_{W})$ is sampled directly from its full conditional distribution. First note that, conditional on $(T,V_T,\theta)$, $(U,W,V_{W})$ is independent of the data and consists of the virtual jumps. The full conditional distribution of $(U,W,V_{W})$ is given by Proposition \ref{prop_2} below.
\begin{proposition}\label{prop_2}
Defining $U^{(i)}$ as the virtual jumps in $(T_i,T_{i+1})$, with $T_0=0$ and $T_{|T|+1}=S$, the full conditional distribution of the virtual jumps is such that:
\begin{enumerate}[i.]
  \item the $U^{(i)}$'s are mutually independent;
  \item for $i=0,\ldots,|T|$, $U^{(i)}$ is a homogeneous Poisson process with rate $\Omega_{h(V_i)}+Q_{h(V_i)}$, where $V_i$ is the state of $V$ at $T_i$.
\end{enumerate}
\end{proposition}
\begin{proof}
See Appendix A.
\end{proof}

Note that, if $\Omega_k = |Q_k|$, the number of virtual jumps is a.s. zero and, as a consequence, the MCMC chain is not irreducible since the non-virtual jumps would be restricted to the set defined by its initial value. In fact, the values of the $\Omega_k$'s have great impact on the efficiency of the algorithm. If these are increased, the mean number of virtual jumps also does which, in turn, improves the mixing of the chain. On the other hand, an increase in the number of virtual jumps leads to an increase in the computational cost of the algorithm, in particular, on the step where $(V_0,V,U,T,R)$ is sampled. \citet{rao2013fast} suggests the use of $\Omega = 2\max\limits_k|Q_k|$ in the context of inference for MMPP, based on empirical results. Note however that the authors consider a unique $\Omega$ for all $k$, as presented in the original augmented CTMC representation of \citet{hobolth2009simulation}. This leads to different local mixing properties of the MCMC with respect to different states (in our case, different functional forms) of the CTMC. Moreover, an optimal local choice w.r.t. the state (functional form) with the larger $|Q_k|$ ought to penalize the local computational cost associated to the other states. That issue is the main motivation for us to propose the alternative augmented CTMC with distinguished $\Omega_k$'s. It allows for a finer optimization of the chain's properties in the sense of globally optimizing the mixing without penalizing the computational cost. Finally, based on the results of \citet{rao2013fast}, we set $\Omega_k = 2|Q_k|$ for all $k$.

\subsubsection*{Sampling $(V_0,V,U,T,R)$}

The density in (\ref{eq_post_dens}) implies that
\begin{align*}
\pi(V_0,V,&U,T,R|y,\theta,\psi,W) \propto L_{0}(V_0,R_0)\pi_0(V_0)\nonumber\\
&\times\prod_{l=1}^{|W|}L_{l}(V_{0:l},R_{0:(l)})\pi(V_l|V_{l-1}, \theta)\\
&\times\prod_{i=0}^{(l)}\pi(R_{(l)}|V_{(l)}),
\end{align*}
where $V_{0:l}=(V_0,\ldots,V_l)$, $(l)$ is the number of non-virtual jumps up to $W_l$ and $R_{0:(l)}$ are all the starting values up to $W_l$ (which are not necessarily $l$ values). Also,
\begin{eqnarray*}
L_l(V_{0:l},R_{0:(l)}) &=& \exp \left\lbrace - \int\limits_{W_l}^{W_{l+1}} \lambda(s)ds \right\rbrace\\
  &&\times\prod\limits_{t_n \in [W_l, W_{l+1})} \lambda(t_n).
  \end{eqnarray*}

Directly sampling from the full conditional distribution of $(V_0,V,U,T,R)$ requires the computation of its probability mass function, which is a (at least) $E^{|W|}$-dimensional vector. Therefore, in the majority of cases, the computational cost associated to this algorithm is impractical. Furthermore, note that, for GMMPP's, each likelihood term $L_l(V_{0:l},R_{0:(l)})$ depends on $V$ and $R$ up to time $W_l$, because of the dependence on the last non-virtual jump up to $W_l$. For that reason, unlike in the case of inference for MMPP's \citep[see][]{rao2013fast}, a FFBS scheme cannot be devised to sample from the full conditional distribution of $(V_0,V,U,T,R)$. Instead, we propose an independent Metropolis Hastings (MH) step with a proposal distribution $q(V_0,V,R)$ that aims at approximating the target full conditional by adding suitable normalizing constant terms for each of the $|W|$ terms $L_{l}(V_{0:l},R_{0:(l)})\pi(V_l|V_{l-1}, \theta)$ or $L_{l}(V_{0:l},R_{0:(l)})\pi(V_l|V_{l-1}, \theta)\pi(R_{(l)}|V_{(l)})$, accordingly. More specifically,
\begin{align}\label{eq_propMH}
&q(V_0,V,R) = c_0 \frac{\pi_0(V_0)}{c_0(V_0)} c_0(V_0) L_0(V_0,R_0)  \\
&\times \pi(R_0 | V_0, \tau) \prod_{l=1}^{|W|} \biggl{\{}c_l \frac{\pi(V_l | V_{l-1}, \theta)}{c_l(V_l)} c_l(V_l)  \nonumber\\
&\times L_l(V_{0:l},R_{0:(l)})[\pi(R_{(l)}|V_l)\I(V_{l-1} \neq V_l) \nonumber\\
&+ \I(V_{l-1} = V_l)]\biggr{\}},\nonumber
\end{align}
where $c_l(V_l)$ is the normalizing constant of $L_l(V_{0:l},R_{0:(l)})\pi(R_{(l)}|V_l)$ and $c_l$ is the normalizing constant of $\ds\frac{\pi(V_l|V_{l-1},\theta)}{c_l(V_l)}$. Note that when $V_l$ corresponds to a virtual jump, $c_l$ is the constant that normalizes $\pi(V_l|V_{l-1},\theta)L_l(V_{0:l},R_{0:(l)})$.

The acceptance probability of the MH step is given by
\begin{equation}\label{eq_apMH}
\alpha = 1 \wedge \prod_{l=1}^{|W|} \frac{c_l}{c_{l}^*},
\end{equation}
where the $c_l$'s and $c_{l}^*$'s refer to the current and proposal values, respectively. Note that any trajectory that leads negative values for the IF is rejected with probability 1.

As describe before, the constant functional form is the only one for which we assume a continuous prior for its starting values - a Gamma prior. For all the other forms, the required normalizing constants above would typically be intractable for continuous priors. The detailed algorithm to perform the MH step described above is presented in Algorithm \ref{alg_1} of Appendix B. The algorithm for the simpler case with no varying starting value is obtained by applying the straightforward simplifications.

We use the general result from \citet{mengersen1996rates} to establish the uniform ergodicity of the proposed MH sub-chain.
\begin{proposition}\label{prop_3}
The Metropolis-Hastings sub-chain defined by (\ref{eq_propMH}) and (\ref{eq_apMH}) is uniformly ergodic.
\end{proposition}
\begin{proof}
See Appendix A.
\end{proof}

Since this is an independent MH algorithm, its efficiency relies heavily on its acceptance rate - the higher the better. Now note that this rate ought to reduce as the number of non-virtual jumps increases. For that reason, we propose an adaptation of the algorithm above that partitions the interval $[0,S]$ and separately samples $(V_0,V,U,T,R)$ in each of these time intervals from its respective full conditional distribution. In order to have a robustly efficient algorithm, we propose an adapting strategy that starts by updating $(V_0,V,U,T,R)$ in one block and then partitions this into more blocks if required. The adaptation is considered up to a certain iteration of the Markov chain so to guarantee its convergence. Finally, the algorithm to sample $(V_0,V,U,T,R)$ in each sub-interval of time is a direct and straightforward adaptation of Algorithm \ref{alg_1}.

The partitioning strategy ought to be executed with care in order to guarantee that the respective full conditional distribution depends on the likelihood only inside the respective time interval. This means that the limits of the intervals have to be times $T_i$'s of non-virtual jumps and the proposal distribution requires the restriction that preservers the upper limit of the respective time interval as a non-virtual jump time. That is achieved as follows.

Any blocking scheme based on a partition $(0=s_0,s_1,\ldots,s_B=S)$ must be such that, for the current state of $(V_0,V,U,T,R)$, times $s_b$, $b=1,\ldots,B-1$ are non-virtual jump times in which the CTMC assumes values among the first $K$ states in its state space. Furthermore, the blocks are defined by the intervals $[0,s_1)$, $[s_1,s_2)$, $\ldots$, $[s_{B-2},s_{B-1})$, $[s_{B-1},S]$.

The adapting partition strategy goes as follows. Set a number of iterations $M$ large enough to obtain reliable estimations of the acceptance rate and a reasonable threshold $r$ for the rate. The algorithm starts with one block $(B=1)$. Then, after every $M$ iterations, the acceptance rate in those last $M$ iterations is evaluated. If this rate is smaller than $r$, we make $B=B+1$. The adaptation carries on until the computed rate is larger than $r$. We suggest $r\approx0.25$. Finally, a partition with $B$ blocks is defined by setting the intervals' limits to be the $T_i$'s which are the closest to the times $|W|/B,\;2|W|/B,\ldots,\;(B-1)|W|/B$.

Finally, continuous time Markov chain trajectory may be highly correlated to some parameters in $\psi$, which may compromise the mixing of those parameters. A simple and efficient way to mitigate this problem is to perform multiple updates of $(U,W,V_{W})$ and $(V_0,V,U,T,R)$ on each iteration of the Gibbs sampling. This issue is illustrate in the simulated examples.

\subsubsection*{Sampling $\theta$ and $\psi$}

It is straightforward to simulate from the full conditional distribution of $\theta$ given the conditional independence structure and conjugation of its prior. The full conditional distribution of $\theta_k$ is a  $Gamma(\alpha_k+m_k(V_0,V),\beta_k+\tau_k(V_0,V,T))$,
where $m_k(V_0,V)$ and $\tau_k(V_0,V,T)$ are the total number of visits to and the total time spent at the $k$-th functional form in $[0,S]$, respectively. Moreover $\theta_{k \cdot}$ has a Dirichlet full conditional distribution with parameter vector
$(\gamma_{k1}+m_{k1}(V_0,V),\ldots,\gamma_{kK}+m_{kK}(V_0,V))$,
if self-jumps are allowed, and
$(...,\gamma_{kk-1}+m_{kk-1}(V_0,V),\gamma_{kk+1}+m_{kk+1}(V_0,V),...)$,
if no self-jumps are allowed, where $m_{k_1k_2}(V_0,V)$ is the total number of transitions from functional form $k_1$ to functional form $k_2$ in $[0,S]$,

Concerning parameters $\psi$, we have from (\ref{eq_post_dens}) that
\begin{equation*}
\pi(\psi)\propto \exp\left\{-\int_{0}^S\lambda(s)ds\right\}\prod_{n=1}^{N_S}\lambda(t_n)\pi(\psi).
\end{equation*}
The prior independence among the $\psi_k$'s implies in the conditional independence of the respective $K$ full conditionals. Moreover, for a constant $g_k$ with fixed starting value, a
$Gamma(\eta_k,\nu_k)$ leads to a full conditional
\begin{equation*}
Gamma(\eta_k+n_k(y),\nu_k+\tau_k(V_0,V,T)),
\end{equation*}
where $n_k(y)$ is the number of events from $y$ occurring during the time that the IF assumes the functional form $g_k$. For all the other functional forms we perform MH steps with an adapted random walk proposal \citep[see][Section 2]{roberts2009examples} for each vector $\psi_k$. The acceptance probability for each $\psi_k$ is given by
$$1\wedge\frac{\pi(\psi_{k}^*|\cdot)}{\pi(\psi_k|\cdot)},$$
where $\psi_k$ and $\psi_{k}^*$ are the current and proposal values, respectively, and $\pi(\psi_k|\cdot)$ is proportional to the product of the likelihood in (\ref{eq_likelihood}) and the prior density of $\psi_k$.

\subsection{Prediction}

Prediction is a common procedure associated to the statistical analysis of stochastic processes. In the context of unidimensional Poisson processes, prediction consists in estimating the future behavior of the process, in particular, its intensity function and/or events. The Bayesian approach allows prediction to be made under a probabilistic approach through the predictive distribution. Consider the full Bayesian model of a GMMPP $Y$ in $[0,\infty]$ and let $y$ be a realization of the process in $[0,S]$. Now define $g(Y,V,T,\psi)$ to be some measurable function, in the probability space of the full Bayesian model, that depends on $(Y,V,T)$ only in $(S,\infty)$. Then, prediction about $g(Y,V,T,\psi)$ is made through the predictive distribution of $g(Y,V,T,\psi)|y$.

In a MCMC context, it is straightforward to obtain a Monte Carlo (MC) sample from the predictive distribution as long as it is feasible to simulate from the full model.
A MC sample is obtained by simulating $g(Y,V,T)$ conditional on each value simulated along the MCMC (after a burn-in period) due to the fact that
\begin{align*}
\pi(&g(Y,V,T,\psi)|y)=\\
&\int \pi(g(Y,V,T,\psi)|\varphi,y)\pi(\varphi|y)d\varphi.
\end{align*}

Appealing examples of $g(Y,V,T,\psi)$ include:
\begin{enumerate}[i.]
  \item $\ds\dot{\lambda} := \{\lambda(s)\}_{s \in (S,S+\dot{S})}$, for $\dot{S}>0$;
  \item $\ds \Lambda_{\dot{S}}=\int_{S}^{S+\dot{S}}\lambda(s)ds$;
  \item $N_{S+\dot{S}}-N_S$.
\end{enumerate}
For examples $i.$ and $ii.$, it is enough to simulate the CTMC $X$ conditional on each sample of $(X_S,\theta)$ and compute $g$ for the respective sampled value of $\psi$. For example $iii.$, an extra step is required to simulate from a $Poisson(\Lambda_{\dot{S}})$ distribution, conditional on each simulated valued of $\Lambda_{\dot{S}}$.


\section{Simulated examples}\label{sec.sim}

This section presents a collection of simulated examples to explore important issues related to the methodology proposed in this paper. In particular, we explore: 1. the impact of the number of observations and the number of jumps in the IF on the computational cost of the MCMC algorithm; 2. a sensitivity analysis for the priors of $\psi$ and $\theta$; 3. the efficiency in estimation and prediction (with replications).

Convergence diagnostics are obtained based on the MCMC chain for the parameters, for some functions of the CTMC and for the log-posterior density. Computational cost is evaluated in terms of the average time (in seconds) to obtain 100 effective samples of the log-posterior density. All the reported computational costs were computed after calibrating the MH proposal for $\psi$. The effective sample size of an MCMC sample of size $n$ is defined as $n_{ess}=\frac{n}{1+2\sum_{j=1}^{\infty}\rho_j}$, where $\rho_j$ is the autocorrelation of order $j$ of the chain. It is such that the variance of the ergodic average of the $n$ values from the chain is the same as the variance of the ergodic average of an independent sample (from the target distribution) of size $n_{ess}$. For the two examples in which the GMMPP is compared to a non-parametric IF model, we consider the effective samples of the log-likelihood instead of the log-posterior density.

All the examples are implemented in Ox \citep{doornik2009object} and run on an i7 3.4GHz processor with 16MB RAM. Codes are available upon request to the authors.

\subsection{Investigating the computational cost}\label{subseccompcost}

In this section we investigate the computational cost associated to the proposed methodology. In particular, we investigate the impact of the number of observations and the number of changes in the IF trajectory. As it has been emphasized before, the low computational cost is at the core of the main contributions of this paper.

\subsubsection*{The cost as a function of the number of observations}

We simulate five scenarios with the same behavior for the IF (functional forms and changes) in the same time interval but with different levels of magnitude. We consider three functional forms - increasing and decreasing lines and a constant, with fixed starting values and no self jumps allowed. Table \ref{tab1} presents the specific functional forms, length of stay, mean number and actual number of observations. We fix the Q-matrix so that all the states have mean staying time of 20 units and uniformly distributed transition probabilities.
Multiple updates of blocks $(U,W,V_{W})$ and $(V_0,V,U,T,R)$ are performed to control the high autocorrelation of the parameters of the increasing line functional form - 5 updates for scenarios A1 and A2, 15 for A3 and A4 and 25 for A5. Moderately informative priors are adopted for $\psi$ in scenario $A1$, namely $\psi_{11}\sim N(1,2^2)$ (intercept of the increasing line),  $\psi_{12}\sim N(0.5,0.6^2)$ (slope of the increasing line), $\psi_{21}\sim N(5,2^2)$ (intercept of the decreasing line), $\psi_{22}\sim N(-0.5,0.6^2)$ (slope of the decreasing line), $\psi_{31}\sim Gamma(1,1)$ (constant). For all the other scenarios, independent uniform improper priors are used for all parameters but $\psi_{31}$, for which a $Gamma(1,1)$ is also used.

Results regarding the estimation of the IF and of the $\psi$ parameters are presented in Figure \ref{fig_ap_1} and Tables \ref{tab_ap_1} and \ref{tab_ap_4} in Appendix C. They show a reasonably good recovery of the IF and parameters already for the data set with only 103 observations with the estimation improving substantially with the size of the data set. MCMC diagnostics are presented in Figures  \ref{fig_ap_9} and \ref{fig_ap_13} in Appendix D.

\begin{figure}[h!]
	\centering{
		\includegraphics[width=0.6\linewidth]{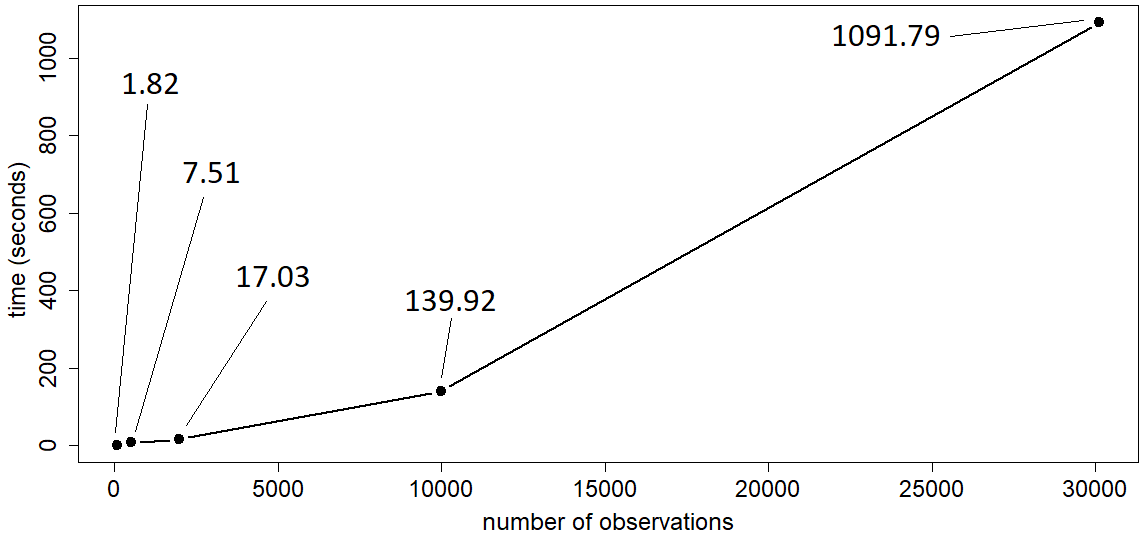}
		\caption{Computational cost, measured in terms of the time per 100 effect samples from the log-posterior density, versus number of observations.}
		\label{fig2}}
\end{figure}

The relation between the computational cost and the number of observations is shown in Figure \ref{fig2}. We highlight the computational efficiency of the proposed MCMC algorithm shown by the running times. The methodology has shown to be quite efficient to be applied for very large data sets. For example, the total running time to obtain an effective sample size of 100 for the log-posterior density is around 2.3 minutes for the data set with 10 thousand observations and 18 minutes for the data set with 30 thousand observations.

\begin{table*}[!h]
  \caption{Description of the five scenarios simulated - $\psi$, length of stay, mean and actual number of observations.}\label{tab1}
  \centering
  \begin{tabular*}{\textwidth}{@{\extracolsep{\fill}}c|c|c|c|c|c|c|c|c|c|c}
    \hline
    scen. & \multicolumn{3}{c|}{$g_2$} & \multicolumn{3}{c|}{$g_1$} & \multicolumn{2}{c|}{$g_3$} & $E[N_S]$ & $N_S$ \\ \hline
          & $\psi_{21}$ & $\psi_{21}$ & stay & $\psi_{11}$ & $\psi_{12}$ & stay & $\psi_{31}$ & stay &  & \\ \hline
  A1 & 6    & -0.25 & [0,14.2) &  0.5 & 0.25 & [36.5,50] & 0.5 & [14.2,36.5) & 100.68 & 103  \\
  A2 & 30   & -1.25 & [0,14.2) &  2.5 & 1.25 & [36.5,50] & 2.5 & [14.2,36.5) & 503.38 & 512  \\
  A3 & 120  & -5    & [0,14.2) &  10  & 5    & [36.5,50] & 10  & [14.2,36.5) & 2013.5 & 1985 \\
  A4 & 600  & -25   & [0,14.2) &  50  & 25   & [36.5,50] & 50  & [14.2,36.5) & 10068  & 9991 \\
  A5 & 1800 & -75   & [0,14.2) &  150 & 75   & [36.5,50] & 150 & [14.2,36.5) & 30203  & 30113\\
    \hline
  \end{tabular*}
\end{table*}

\subsubsection*{The cost as a function of the number of changes in the IF}

We simulate three scenarios with the same average number of observations and three functional forms - increasing and decreasing lines and a constant, with fixed starting values and self jumps allowed. The IF is simulated from the same CTMC prior but considering different total observed time in order to have considerably different numbers of changes in the IF. The number of observations is approximately 2000 for all the scenarios. Table \ref{tab2} presents the specific functional forms, average length of stay per visit and number of changes in the IF. The priors on the Q-matrix diagonal parameters are $\theta_1\sim Gamma(1,10)$, $\theta_2\sim Gamma(1,10)$ and $\theta_3\sim Gamma(1,5)$ and, for the transition probability vectors, we adopt a uniform prior on the respective simplex. Finally, uniform improper priors are adopted for all the $\psi$ parameters. Blocks $(U,W,V_{W})$ and $(V_0,V,U,T,R)$ are updated 5 times in each iteration of the Gibbs sampling.

Results regarding the estimation of the IF and of the $\psi$ and $\theta$ parameters are presented in Figure \ref{fig_ap_2} and Tables \ref{tab_ap_2} and \ref{tab_ap_5} in Appendix C. They show a very good recovery of the IF and parameters. MCMC diagnostics are presented in Figures \ref{fig_ap_14} and \ref{fig_ap_16} in Appendix D.

\begin{figure}[h!]
	\centering{
		\includegraphics[width=0.6\linewidth]{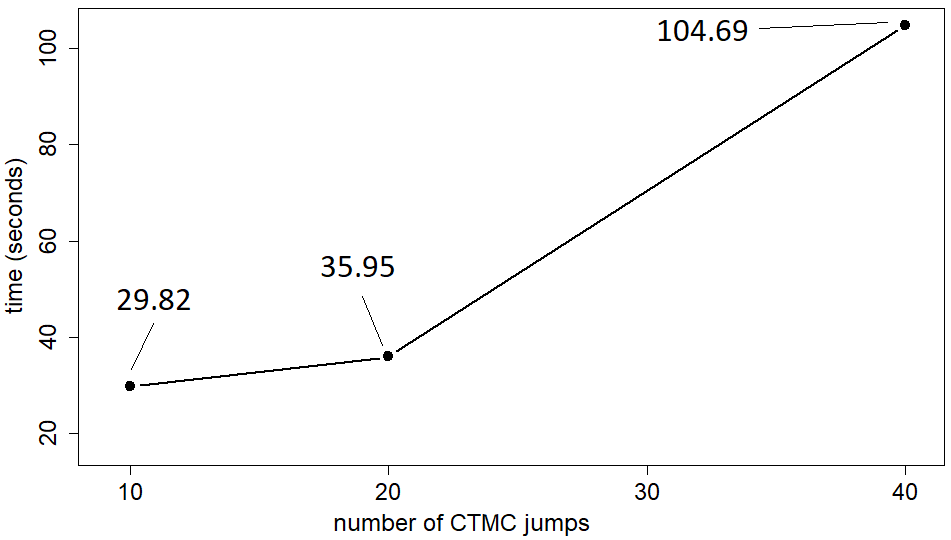}
		\caption{Computational cost, measured in terms of the time per 100 effect samples from the log-posterior density, versus number of changes in the IF.}
		\label{fig3}}
\end{figure}

The relation between the computational cost and the real number of changes in the IF is shown in Figure \ref{fig3}. Again, we highlight the computational efficiency of the proposed MCMC algorithm shown by the running times. The total running time to obtain an effective sample size of 100 for the log-posterior density is around 105 seconds for the data set with 40 changes in the IF and approximately 2 thousand observations.

\begin{table*}[!h]
  \caption{Description of the three scenarios simulated - $\psi$, average length of stay per visit, number of changes in the IF.}\label{tab2}
  \centering
  \begin{tabular*}{\textwidth}{@{\extracolsep{\fill}}c|c|c|c|c|c|c|c|c|c|c}
    \hline
    scen. & \multicolumn{3}{c|}{$g_2$} & \multicolumn{3}{c|}{$g_1$} & \multicolumn{2}{c|}{$g_3$} & $|T|$  & $N_S$ \\ \hline
          & $\psi_{21}$ & $\psi_{22}$ & av. st. &  $\psi_{11}$ & $\psi_{12}$ & av. st. & $\psi_{31}$ & av. st. & &  \\ \hline
        B1 & 20.4 & -0.85 & 15 &  1.7 & 0.85 & 15 & 1.7 & 15 & 10 & 2013 \\
        B2 & 12   & -0.5  & 15 &  1   & 0.5  & 15 & 1   & 15 & 20 & 2004 \\
        B3 & 6    & -0.25 & 15 &  0.5 & 0.25 & 15 & 0.5 & 15 & 40 & 1991 \\
    \hline
  \end{tabular*}
\end{table*}

\subsection{Prior sensitivity analysis}\label{subsecpriorsenst}

We perform a prior sensitivity analysis for parameters $\psi$ for scenarios A1, A3, A5 and B2. Those examples are run with non-informative and moderately informative priors. The latter are set based on the scale of each example. Also, a prior sensitive analysis for parameters $\theta$ in the diagonal of the Q-matrix is performed for scenarios B1 and B3. Again, non-informative and moderately informative priors are used.

In the first analysis, the Q-matrix is fixed for all the A$*$ scenarios in the same values as in Section \ref{subseccompcost}. For scenario B2, the same non-informative priors from Section \ref{subseccompcost} are adopted. The prior on the constant IF parameter $\psi_{31}$ is set to be $Gamma(1,1)$ in all the cases. For the parameters indexing the other two functional forms, we compare the results for improper uniform priors and the moderately informative priors shown in Table \ref{tab3}. Results for the parameters estimation are shown in Table \ref{tab_ap_3} in Appendix C and show that greater differences are observed only for the parameters of the increasing line. As it should be expected, the variances of those parameters are greater for the non-informative priors, for which the posterior density is also more asymmetric. Results for the IF go in the same direction, with significant differences observed only for scenario A1 in the time period associated to the increasing line. It can be noticed that the posterior distribution of the IF is more influenced by the data for the non-informative prior, as expected - see Figure \ref{fig_ap_3} in Appendix C.

\begin{table*}[!h]
  \caption{Informative priors for the sensitivity analysis of $\psi$.}\label{tab3}
  \centering
  \begin{tabular*}{\textwidth}{@{\extracolsep{\fill}}c|c|c|c|c}
    \hline
    scen. & $\psi_{11}$    & $\psi_{12}$    & $\psi_{21}$     & $\psi_{22}$ \\ \hline
     A1   &  $N(1,2^2)$    & $N(0.5,0.6^2)$ & $N(5,2^2)$      & $N(-0.5,0.6^2)$ \\
     A3   &  $N(10,5^2)$   & $N(6,4^2)$     & $N(120,20^2)$   & $N(-6,4^2)$ \\
     A5   &  $N(150,50^2)$ & $N(75,40^2)$   & $N(1800,150^2)$ & $N(-75,40^2)$ \\
     B2   &  $N(2,2^2)$    & $N(1,1)$       & $N(12,4^2)$     & $N(-1,1)$ \\
    \hline
  \end{tabular*}
\end{table*}

The second sensitivity analysis concerns the prior distribution on the parameters $\theta$ in the diagonal of the Q-matrix. Improper uniform priors are adopted for all the $\psi$ parameters except for the constant value $\psi_{31}$ which has a $Gamma(1,1)$ non-informative prior. Uniform priors on the simplex are adopted for all the transition probability vectors in the Q-matrix. The non-informative priors for the Q-matrix diagonal parameters are improper uniforms distributions and the informative ones are $\theta_1\sim Gamma(1,10)$, $\theta_2\sim Gamma(1,10)$ and $\theta_3\sim Gamma(1,5)$. Results (omitted here) are virtually the same for the two prior specifications w.r.t. the estimated IF, $\psi$ parameters and transition probability vectors from the Q-matrix. As for the parameters in the diagonal of the Q-matrix, small yet non-negligible differences are observed, with slightly larger variances for the non-informative priors case.

\subsection{Efficiency in estimation and prediction}

\subsubsection*{Examples with replications}

We now investigate the efficiency of the proposed methodology in terms of estimation and prediction by considering replications of the same model. We consider the IF from scenarios A1, A3 and B2 and generate 50 independent data sets for each one. Prediction for the integrated IF in interval $[400,800]$ is performed for scenario B2 by sampling from its predictive distribution.

In order to summarize the performance of the proposed model we consider the posterior distribution of the following measure of fit:
\begin{equation}\label{meas_fit}
\frac{1}{S} \int_0^S|\lambda(s) -\lambda_R(s)|ds,
\end{equation}
where $\lambda_R$ is the real intensity function.

Results are shown in Figures \ref{fig_ap_4}, \ref{fig_ap_5} and \ref{fig_ap_6} and reveal a very good performance of the proposed methodology to estimate and predict the IF.

\subsubsection*{Comparison to non-parametric Cox process model}

We compare the proposed class of models to that proposed in \citet{G&G}, in which the IF is assumed to be a continuous positive function of a latent Gaussian process. The computational cost associated to the MCMC algorithm from \citet{G&G} is $\mathcal{O}((\lambda_{sup}S)^3)$, where $\lambda_{sup}$ is the supremum of the IF in $[0,S]$. It is defined by the cost to generate multivariate normal distributions which are required due to the use of a latent Gaussian process. This implies that not only the cost is larger and grows much faster than the cost from our methodology but also that it is not feasible to apply the methodology to very large data sets.

We consider a data set of size 302 generated from the IF $\lambda(s)=20\exp\{-x/5\}+1.5\exp\{-(x-25)^2/50\}$ in $[0,50]$ and excluding the 8 observations generated in $[9,14]$. The GMMPP is fit with decreasing and increasing lines and a constant. The estimated IF for both models are shown in Figure \ref{fig_ap_7} in Appendix C and show that similar results are obtained for both models. The computational time per 100 effective samples of the log-likelihood is 15.32 seconds for the GMMPP and 2 hours for the non-parametric IF model (with no approximations to simulate from the Gaussian process).

\section{Applications}\label{sec.app}

\subsection{Coal mining disasters}

We apply the proposed methodology to the classic coal mining disasters data of \citet{jarret79}, consisting of the dates of 191 explosions in coal mines that killed ten or more men in Britain between 15th March 1851 and 22th March 1962 (re-scaled to $[0,112]$, year unit). We also analyze data with the non-parametric IF model of \citet{G&G}.

Based on an empirical analysis of the data (see Figure \ref{fig0}), we set two functional forms - a decreasing and an increasing line with varying starting values. We adopt the following priors: $\mbox{Uniform}(0.1,0.3,\ldots,4.7)$ for both the varying starting values, improper uniform for both the slopes, $Gamma(1,20)$ for all the diagonal parameters of the Q-matrix and uniform priors are used for the transition probabilities.

\begin{figure*}[h!]
	\centering{
		\includegraphics[width=0.8\linewidth]{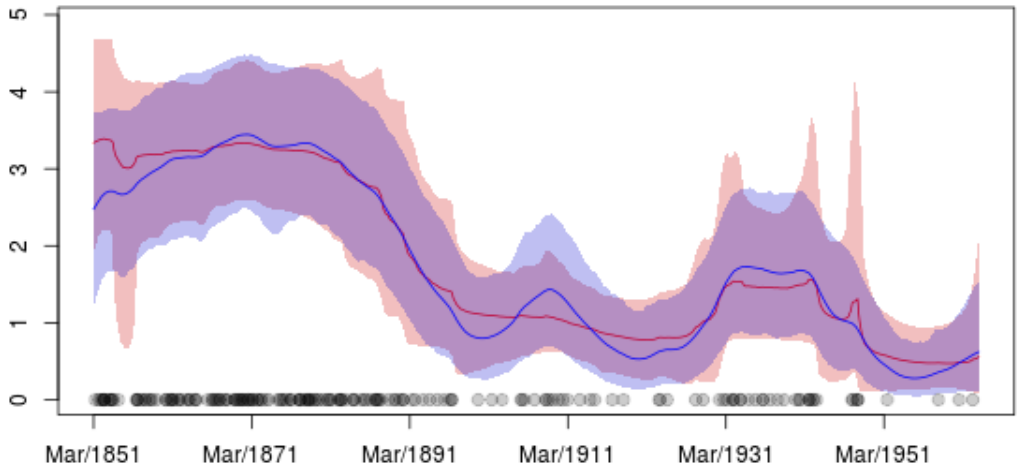}
		\caption{Estimated IF (posterior mean and 95\% CI) for the coal mining data for the GMMPP (red) and non-parametric IF (blue) models.}
		\label{fig4}}
\end{figure*}

The computational time per 100 effective samples of the log-likelihood is 2.93 minutes for the GMMPP and 8.6 minutes for the non-parametric IF model (with no approximations to simulate from the Gaussian process). The estimated IF for both models is shown in Figure \ref{fig4}. The posterior mean and standard deviation of the integrated IF is 197.6 and 13.8 for the GMMPP and 193.4 and 14.5 for the non-parametric IF model. The mean and standard deviation of the slopes are -0.0406 and 0.0388 for the negative one and 0.1620 and 0.3709 for the positive one. The same statistics for the mean waiting times are 11.70 and 4.18 for the decreasing line and 9.36 and 3.16 for the increasing line.

\subsection{BRLxUSD exchange rate}

We consider the exchange rate between US Dollar to Brazilian Real. The data set consists of the 1163 days, between Jan 2000 and Dec 2017 (re-scaled to $[0,216]$, month unit), in which the exchange rate varied more than 1$\%$. Prediction is performed for the period of Jan 2018 to Apr 2020.

Based on an empirical analysis of the data (see Figure \ref{fig0}), we set two functional forms - a decreasing and an increasing lines, both with varying starting values. We adopt the following priors:  $\mbox{Uniform}(5,5.2,\ldots,12)$ and  $\mbox{Uniform}(0,0.2,\ldots,6)$ for the varying starting values of the decreasing and increasing lines, respectively, and uniform improper priors for both the slopes. A $Gamma(1,40)$ is assumed for the diagonal parameters of the Q-matrix and uniform priors are used for the transition probabilities. The observed time interval is divided into 6 blocks to update the CTMC component.

\begin{figure*}[h!]
	\centering{
		\includegraphics[width=0.8\linewidth]{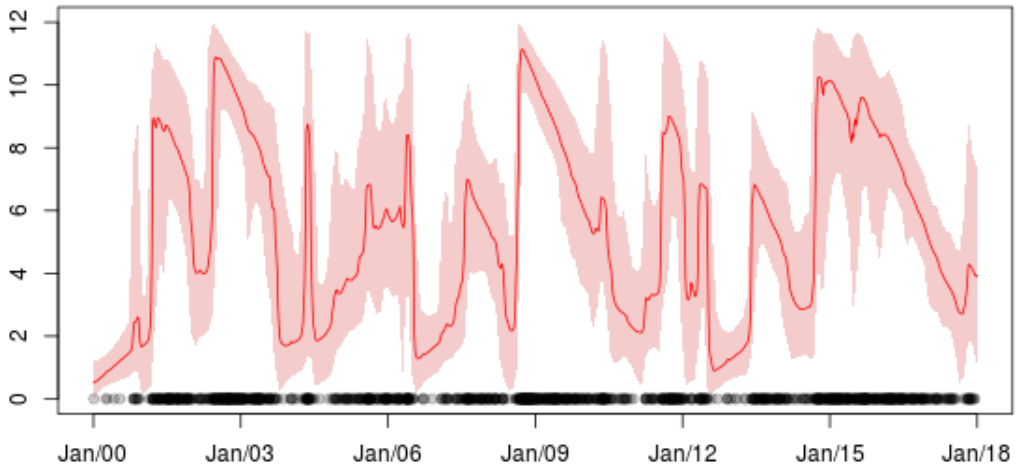}
		\caption{Estimated IF (posterior mean and 95\% CI) for the exchange rate data.}
		\label{fig5}}
\end{figure*}

The MCMC algorithm takes around 4.6 minutes to draw 100 effective samples. The estimated IF is shown in Figure \ref{fig5}. The posterior mean and standard deviation of the integrated IF is 1143.1 and 34.3. The mean and standard deviation of the slopes are \text{--}0.3102 and 0.0651 for the negative one and 0.1176 and 0.0355 for the positive one. The same statistics for the mean waiting times are 11.70 and 4.18 for the decreasing line and 9.36 and 3.16 for the increasing line.

The predictive distribution of the integrated intensity between Jan 2018 to Apr 2020 is shown in Figure \ref{fig_ap_8} in Appendix C and has mean 151.4, standard deviation 35.80 and 95\% CI $(78.44,218.29)$. The real observed number of events is 151.

\subsection{Modeling epidemic curves}

We consider a model which we believe to be of practical use to model epidemic phenomena. The idea is to model each cycle of the IF to have an exponential growth, some period of stabilization and then an exponential decay. Moreover, in order to mimic the expected behavior of epidemic curves, we need the exponential growth and decay rates to change over time. This behavior can be emulated by using a cdf, in particular the standard normal cdf. The parametrization of the model is such that the model is flexible and parameters have a clear interpretation. The idea is to model each cycle of the epidemic phenomena after the IF starts to decrease. This means that the IF is known to start in the increasing functional form and than having one change to the decreasing one and, therefore, prediction would concern its future decreasing behavior. The model is the following:
\begin{eqnarray*}
 g_1(s,\psi) &=& b_1+a\Phi(d_1+c_1s)\\
 g_2(s,\psi) &=& b_2+\gamma(\psi,T_1)\Phi(d_2-c_2(s-T_1)),
\end{eqnarray*}
where $g_1$ and $g_2$ are the increasing and decreasing curves, respectively, and $\gamma(\psi,T_1)$ is set to be $(b_1+a\Phi(d_1+c_1T_1)-b_2)/\Phi(d_2)$ to guarantee the continuity of the IF at the change time.

We impose some restrictions on the parameter space so to ease model identifiability and parameter interpretation. We set $c_1>0$, $c_2>0$, $a>0$, $b_1\geq0$ and $b_2\geq0$ as standard identifiability restrictions. We also set the less obvious restriction $d_2<3$ so that the time period of constant behavior of the IF is majorally accommodated by the end of the increasing function $g_1$ and, consequently, identifiability of the change point is favored. Also, note that in order to estimate the stabilization level $b_2$ the data needs to include the stabilization period, otherwise, this parameter should be fixed (for example, at zero).

The restrictions above lead to a clear interpretation of the model's parameters as follows.
\begin{itemize}
\item $b_1$: identifies the initial value of the IF - typically around $b_1$;
\item $a$: defines the maximum value assumed by the IF - typically $\approx b_1+a$;
\item $d_1$: defines the initial growth rate of the IF curve;
\item $c_1$: defines the rate in which the growth curve changes and the maximum growth rate;
\item $d_2$: defines the initial decay rate of the IF curve;
\item $c_2$: defines the rate in which the decay curve changes and the maximum decay rate;
\item $b_2$: defines the stabilization level after the epidemic period.
\end{itemize}
Furthermore, the maximum slope of the growth and decay curves are given by $\dot{c}_1=\frac{a}{\sqrt{2\pi}}c_1$ and $\dot{c}_2=\frac{-\gamma(\psi,T_1)}{\sqrt{2\pi}}c_2$, respectively. In order to improve the mixing of the MCMC algorithm by reducing the correlation among parameters, we reparametrize the model in terms of $(\dot{c}_1,\dot{c}_2)$ instead of $(c_1,c_2)$. This implies that
\begin{eqnarray}
 g_1(s,\psi) &=& b_1+a\Phi\left(d_1+\frac{\sqrt{2\pi}}{a}\dot{c}_1s\right), \label{eq_exep3}\\
 g_2(s,\psi) &=& b_2+\gamma(\psi,T_1)\nonumber\\
 &\times& \Phi\left(d_2-\frac{\sqrt{2\pi}}{\gamma(\psi,T_1)}\dot{c}_2(s-T_1)\right), \label{eq_exep4} \\
 \gamma(\psi,T_1)&=&\frac{(b_1+a\Phi(d_1+\frac{\sqrt{2\pi}}{a}\dot{c}_1T_1)-b_2)}{\Phi(d_2)}. \label{eq_exep5}
\end{eqnarray}

We highlight the fact that the Bayesian approach and the variance of the Poisson process conditioned on its IF provide considerable flexibility and suitable uncertainty quantification to model epidemic curves, especially when compared to deterministic models directly applied to the number of events.

Finally, this model can be extended to have more flexible curves by considering the cdf of other distributions such as student-t, skew-normal and skew-t. This can account, for example, for skewed growth and decay curves and for cases in which the epidemic curve decays faster than it grows up to a certain time but then takes longer to stabilize, suggesting the use of a heavy tail cdf to model the decay.

\subsubsection{Inference for the epidemic model}

The model in (\ref{eq_exep3})-(\ref{eq_exep5}) has features that allow for some improvements in the MCMC from Section \ref{ssec.mcmc}. It is now possible to sample directly from the full conditional distribution of the block $(V_0,V,U,T,R)$ at a reasonable computational cost. That is because the condition of having only one change in the IF is imposed and, therefore, the size of the state space of this discrete full conditional distribution is $|W|$. This also allows us to increase the value of $\Omega_k$ and, consequently, improve the mixing of the chain, without compromising the cost. One may consider, for example, $\Omega_k=5|Q_k|$.

Another strategy to boost computational efficiency is to truncate the change time to be inside a suitable interval, based on the empirical IF. This interval is conservatively chosen so that it is certain that the change occurs inside it.

\subsubsection{Dengue Fever epidemic}

We analyze data from the 2019 Dengue Fever epidemic in Ceara (CE) state, Brazil, and the 2019/2020 Dengue Fever epidemic in Parana (PR) state, Brazil. The raw data consists of the number of cases per epidemic week, from week 52 of 2018  (23/12/2018) to week 52 of 2019 (28/12/2019) - 371 days, for Ceara, and from week 36 of 2019 (01/09/2018) to week 22 of 2020 (30/05/2020) - 273 days, for Parana. In order to analyze the data, we distribute the cases uniformly in their respective week and use day as the scale unit. The total number of cases in that period was 30700 in Ceara and 331411 in Parana. The epidemic curve in Parana is relatively close to stabilization, but still decaying, so we also predict the time until stabilization - when the IF hits 110. Data is available in the InfoDengue system \citep{infodengue}.

\begin{figure*}[h!]
	\centering{
		\includegraphics[width=1\linewidth]{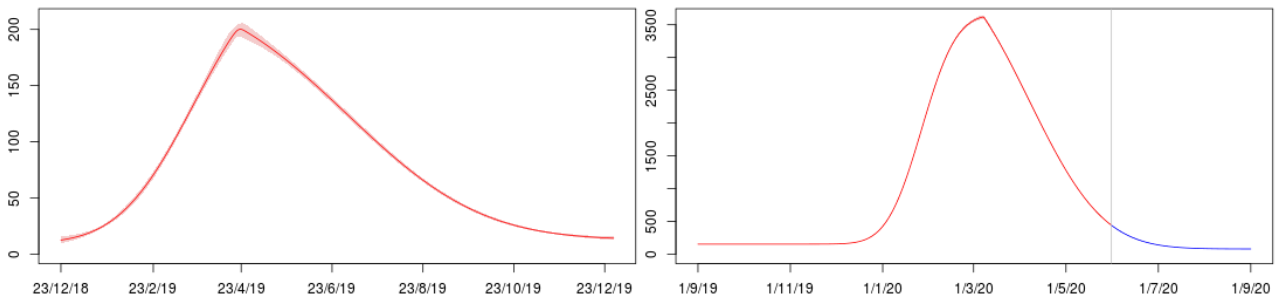}
		\caption{Estimated IF (posterior mean and 95\% CI (red) and predictive mean and 95\% predictive interval (blue)) for the Dengue Fever data - Ceara (left) and Parana (right).}
		\label{fig6}}
\end{figure*}

For the Ceara data, we were compelled to restrict parameter $a$ to be in the interval $(0,250)$ - safely higher than the maximum of the empirical IF, in order to avoid the growth curve to be fit by only (around) half of the cdf. This avoids identifiability and computational problems. For parameter $d_2$ we set priors Uniform$(-\infty,3)$ for Ceara and Uniform$(0,3)$ for Parana. Uniform improper priors are adopted for all the other parameters. The diagonal values of the Q-matrix are fixed at $(1/120,1/250)$, for Ceara, and $(1/180,1/90)$, for Parana. Results are shown in Figure \ref{fig6} and Table \ref{tab4}.

\begin{table*}[!h]
  \caption{Posterior statistics for the Dengue Fever models. IIF is the integrated IF in the observed interval, Pred. time is the predictive time at which the IF reaches 110 and Pred. IIF is the predictive integrated IF until that time.}\label{tab4}
  \centering {\scriptsize
  \begin{tabular*}{\textwidth}{@{\extracolsep{\fill}}c|c|c|c|c|c|c|c|c|c|c|c}
    \hline
       &      &  $a$ & $b_1$ & $c_1$ & $d_1$ & $b_2$ & $c_1$ & $d_2$ & IIF & Pred. Time & Pred. IIF \\ \hline
CE  & mean & 248.3 & 8.80 & 2.47  & -2.22 & 12.78 & 1.25  & 0.99  & 30558.3 & - & - \\
       & s.d. & 1.67  & 2.04 & 0.082 & 0.088 & 1.00  & 0.024 & 0.13  & 174.2 & - & - \\ \hline
PR & mean & 3510.4 & 154.2 & 77.5 & -8.17 & 80 & 46.6 & 0.785 & 331541 & 314.4 (Jul 12$^{th}$) & 9266.8 \\
       & s.d. & 18.9 & 1.28 & 0.66 & 0.087 & - & 0.282 & 0.043 & 574.7 & 0.80 & 242.0\\   \hline
  \end{tabular*}}
\end{table*}

\subsubsection{COVID-19 epidemic}

We also analyze data from the Covid-19 pandemic in Switzerland and Romania. Whilst the epidemic curve has already stabilized for the former, it is still decaying for the latter. For that reason, we predict the time until stabilization - when the IF hits 20, for Romenia. The data set for Switzerland concerns the 30845 cases notified from Feb 25th (date of the first notification) to May 30th - 96 days. The data set from Romania concerns the 19133 cases notified from Feb 26th (date of the first notification) to May 30th - 95 days. Data is obtained from the Coronavirus Resource Center of John Hopkins University through the R package \emph{covid19br} \citep{covid19br}.

\begin{figure*}[h!]
	\centering{
		\includegraphics[width=1\linewidth]{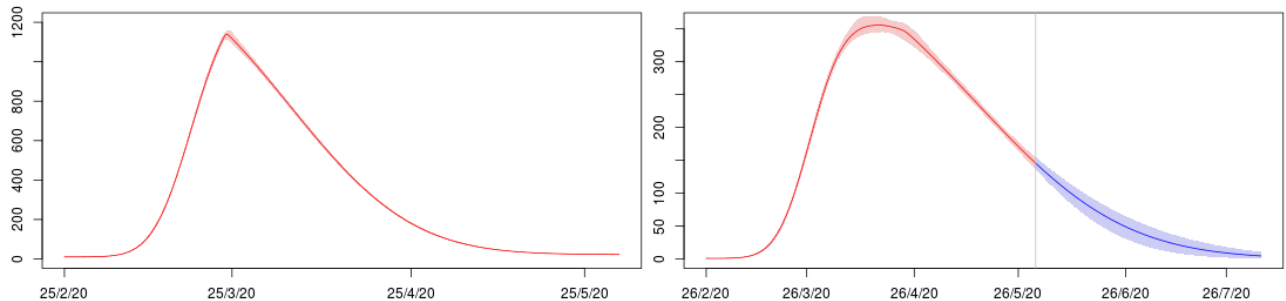}
		\caption{Estimated IF (posterior mean and 95\% CI (red) and predictive mean and 95\% predictive interval (blue)) for the COVID-19 data - Switzerland (left) and Romania (right).}
		\label{fig7}}
\end{figure*}

For the Switzerland data, we restrict parameter $a$ to be in the interval $(0,1300)$ for the same reasons we restrict that parameter for the Ceara Dengue Fever data. For parameter $d_2$ we set priors Uniform$(-\infty,3)$ for Switzerland and Uniform$(0,3)$ for Romania. Uniform improper priors are adopted for all the other parameters. The diagonal values of the Q-matrix are fixed at $(1/30,1/60)$, for Switzerland, and $(1/50,1/40)$, for Romania. Results are shown in Figure \ref{fig7} and Table \ref{tab5}.

\begin{table*}[!h]
  \caption{Posterior statistics for the COVID-19 models. IIF is the integrated IF in the observed interval, Pred. time is the predictive time at which the IF reaches 10 and Pred. IIF is the predictive integrated IF until that time.}\label{tab5}
  \centering {\scriptsize
  \begin{tabular*}{\textwidth}{@{\extracolsep{\fill}}c|c|c|c|c|c|c|c|c|c|c|c}
    \hline
            &      &  $a$   & $b_1$ & $c_1$ & $d_1$ & $b_2$ & $c_1$ & $d_2$ & IIF     & Pred. Time & Pred. IIF \\ \hline
Swi.    & mean & 1295.7 & 10.20 & 98.11 & -4.16 & 22.14 & 36.14 & 0.64  & 30548.9 & - & - \\
            & s.d. & 4.06   & 1.51  & 1.81  & 0.07  & 1.55  & 0.84  & 0.096 & 169.5   & - & - \\ \hline
Rom.     & mean & 358.6  & 0.78  & 19.14 & -4.00 & 0     & 5.77  & 0.66  & 19137.6 & 147.9 (Jul 23$^{rd}$) & 3096.3 \\
            & s.d. & 10.10  & 0.36  & 0.47  & 0.11  & -     & 0.31  & 0.46  & 138.4   & 7.44  & 471.6 \\   \hline
  \end{tabular*}}
\end{table*}

\section{Conclusions}\label{sec.conc}

This paper proposed a novel class of unidimensional Cox processes in which the intensity function assumes predefined functional forms and alternates among these according to the jumps of a continuous time Markov chain. This novel class aims at providing an efficient way to perform useful statistical analysis of unidimensional point processes at a very reasonable computational cost, specially when compared to non-parametric approaches based on latent Gaussian processes.

Important issues regarding model elicitation and identifiability and some aspects of the MCMC algorithm are discussed and explored in simulated studies. Model elicitation should be based on prior knowledge and/or empirical analysis of the data. Whilst non-informative priors work well for the parameters indexing the functional forms, prior elicitation for the parameters in the Q-matrix requires special attention. If not many changes are expected, parameters in the diagonal should be fixed at values coherent with the scale in a way to avoid very short visits. For the transition probabilities, uniform priors are suitable in any case.

The proposed MCMC algorithm performs exact Bayesian inference for the proposed model, so that only Monte Carlo error is involved. The algorithm is carefully devised to efficiently sample from the posterior distribution of all the unknown quantities in the model. In particular, the blocking scheme to sample from the CTMC trajectory has shown to be crucial to obtain a computationally efficient algorithm. Simulated studies illustrated the computational and statistical efficiency of the proposed methodology under different circumstances. In particular, efficient solutions for large data sets are obtained at a reasonable cost.

A particular model to analyze epidemic data is proposed so that asymmetric epidemic curves can be properly accommodated. This model is used to fit data sets regarding Dengue fever in Brazil and COVID-19 in some countries. Results are quite interesting and include the prediction for the curves which have not yet stabilized. The applicability of the methodology for large data sets is illustrated in those examples - one of them have over 300 thousand observations. Other two real data sets are also analyzed to illustrate the applicability of the proposed methodology. Prediction is performed for one of them, providing good results. Finally, results indicate that, typically, models with only straight lines (increasing, decreasing and constant) are enough to provide a good fit.

\section*{Acknowledgements}
The first author would like to thank FAPEMIG and CNPq for financial support. The second author would like to thank CAPES for financial support. The authors would like to thank Fabio Demarqui for helping in obtaining the COVID19 data and Leonardo Bastos for helping in obtaining the Dengue Fever data.

\bibliographystyle{apalike}
\bibliography{biblio}

\section*{Appendix A - Proofs}

\subsection*{Proof of Proposition \ref{prop_1}}

Let $W_{(1)}$ be the first non-virtual jump and $V_{(1)}$ the state of $V$ at $W_{(1)}$. The density of $(W_{(1)}|V_0)$ with respect to the Lebesgue measure is
{\scriptsize\begin{align*}
&\pi_{W_{(1)}|V_0 = j}(t) = \\
&=\sum_{m = 1}^{+\infty} \pi_{W_{(1)}}(t|V_{0:m-1} = j, V_m \neq i)\\
&\textcolor[rgb]{1.00,1.00,1.00}{=\sum_{m = 1}^{+\infty}}\times P(V_{1:m-1} = j, V_m \neq j|V_0 = j) \nonumber \\
&= e^{- \Omega_{h(j)} t} \Omega_{h(j)}(1 - B_{jj}) \sum_{m = 0}^{+\infty}\frac{(\Omega_{h(j)} t B_{jj})^k}{k!} \nonumber \\
&= \Omega_{h(j)} (1 - B_{jj}) e^{- \Omega_{h(j)} (1 - B_{jj})t}\\
&= |Q_{{h(j)}}| e^{-|Q_{{h(j)}}|t} \sim \mbox{Exponential}(|Q_{{h(j)}}|),\; i \in E. \nonumber
\end{align*}}
Similar calculations show that $(W_{l} - W_{l-1}|V_{l-1}=j,W_{l-1}) \sim \mbox{Exponential}(|Q_{h(j)}|)$, $j \in E$, $l \in \mathbb{N}$.
Furthermore,
{\scriptsize\begin{align*}
&P(V_{(1)} = j_2| V_{0} = j_1) = \\
&=\sum_{m=1}^{+\infty} P(V_{m} = j_2|V_{0:m-1} = j_1) P(V_{0:m-1} = j_1|V_0 = j_1) \nonumber \\
&= \sum_{m=1}^{+\infty} B_{j_1j_2}B_{j_1j_1}^{m-1} = B_{j_1j_2}\frac{1}{1 - B_{j_1j_1}}\\
&= -\frac{Q_{j_1j_2}}{Q_{j_1j_1}}, \forall j_2 \neq j_1 \in E.\nonumber
\end{align*}}
Analogous calculations establish the required result for $(V_{l}|V_{l-1})$.
\begin{flushright}
$\square$
\end{flushright}

\subsection*{Proof of Proposition \ref{prop_2}}
Clearly,
{\scriptsize\begin{align*}
&\pi(U,W,V_{W}| T,V_T,\theta) = \frac{\pi(V, T, U, W|\theta)}{\pi(T,V_T|\theta)}\\
&= \frac{\pi_0(V_0) \left[ \prod_{i=0}^{|T|-1} \pi(U^{(i)}, T_{i+1}, V_{i+1}|V_i, T_i)\right] }{\pi_0(V_0) \left[ \prod_{i=0}^{|T|-1} \pi(T_{i+1}, V_{i+1}|V_i, T_i)\right] } \nonumber \\
&\times\frac{\pi(U^{(|T|)}, I_S|V_{|T|}, T_{|T|})}{\pi(I_S|V_{|T|}, T_{|T|})} \nonumber \\
&= \frac{\left[ \prod_{i=0}^{|T|-1} \pi(V_{i+1}|V_i) \pi(T_{i+1}|V_i, T_i) \pi(U^{(i)}|V_i, T_i, T_{i+1})\right] }{\left[ \prod_{i=0}^{|T|-1} \pi(V_{i+1}|V_i) \pi(T_{i+1}|V_i, T_i)\right] \pi(I_S|V_{|T|}, T_{|T|})} \nonumber \\
&\textcolor[rgb]{1.00,1.00,1.00}{=} \times \pi(I_S|V_{|T|}, T_{|T|}) \pi(U^{(|T|)}|V_{|T|}, T_{|T|}, I_S)\\
&= \left[ \prod_{i=0}^{|T|-1} \pi(U^{(i)}|V_i, T_i, T_{i+1})\right] \pi(U^{(|T|)}|V_{|T|}, T_{|T|}, I_S),
\end{align*}}
where $I_S := \I(T_{|T|+1} > S)$. This gives $P(I_S=1|T_{|T|})=e^{-Q_{V_{|T|}}(S-T_{|T|})}$. The result above establishes part $i.$ of the proposition.

We now obtain the full conditional density of $U^{(i)}$, $i=0,\ldots,|T|-1$, w.r.t. the measure of a unit rate Poisson process. We have that
{\scriptsize\begin{align*}
&\pi(U^{(i)}|V_i, T_i, T_{i+1}) = \frac{\pi(U^{(i)}, T_{i+1}|V_i, T_i)}{\pi(T_{i+1}|V_i, T_i)}\\
&= \frac{\Omega_{h(V_i)}^{|U_i|+1}e^{-\Omega_{h(V_i)}(T_{i+1} - T_i)} B_{V_iV_i}^{|U_i|} (1-B_{V_iV_i})}{e^{-(T_{i+1}-T_i)} |Q_{V_i}|e^{-|Q_{V_i}|(T_{i+1}-T_i)} } \nonumber \\
&= \frac{\Omega_{V_i}^{|U_i|+1}e^{-\Omega_{h(V_i)}(T_{i+1} - T_i)} \left( 1 - \frac{ |Q_{V_i}|}{\Omega_{V_i}} \right)^{|U_i|} \left( \frac{|Q_{V_i}|}{\Omega_{h(V_i)}} \right)}{e^{-(T_{i+1}-T_i)} |Q_{V_i}|e^{-|Q_{V_i}|(T_{i+1}-T_i)} }\\
 &= \left( \Omega_{h(V_i)} + Q_{V_i} \right)^{|U_i|} \frac{e^{-\left( \Omega_{h(V_i)} + Q_{V_i} \right)(T_{i+1}-T_i)}}{e^{-(T_{i+1}-T_i)}}, \nonumber
\end{align*}}
where $|U_i|$ is the number of virtual jumps in $[T_i,T_{i+1})$. This establishes part $ii.$ of the proposition for $i=0,\ldots,|T|-1$. For $i=|T|$, we have that
{\scriptsize\begin{align*}
&\pi(U^{(|T|)}|V_{|T|}, T_{|T|}, I_S) = \frac{\pi(U^{(|T|)}, I_S|V_{|T|}, T_{|T|})}{\pi(I_S|V_{|T|}, T_{|T|})}\\
 &= \frac{\Omega_{V_{|T|}}^{|U_{|T|}|}e^{-\Omega_{h(V_{|T|})}(S - T_{|T|})} B_{V_{|T|}V_{|T|}}^{|U_{|T|}|}}{e^{-(S - T_{|T|})} e^{-|Q_{V_{|T|}}|(S-T_{|T|})} } \\
&= \left( \Omega_{h(V_{|T|})} + Q_{V_{|T|}} \right)^{|U_{|T|}|} \frac{e^{-\left( \Omega_{h(V_{|T|})} + Q_{V_{|T|}} \right)(S-T_{|T|})}}{e^{-(S - T_{|T|})}},
\end{align*}}
which concludes the proof.
\begin{flushright}
$\square$
\end{flushright}

\subsection*{Proof of Proposition \ref{prop_3}}

In order to establish uniform ergodicity for an independent MH chain, it is enough to show that the ratio $\displaystyle\frac{q}{\pi}$, where $\pi$ is the target density, is uniformly bounded away from zero on the support of $\pi$ \citep{mengersen1996rates}. Note that
{\scriptsize\begin{equation*}
\frac{q}{\pi}(V_0,V,R) = \frac{1}{\kappa}\prod_{l=1}^{|W|}c_l > \frac{\beta}{\kappa}=\frac{\min_{\mathcal{V}} \prod_{l=0}^{|W|}c_l}{\kappa},
\end{equation*}}
where $\kappa > 0$ is a constant and $\mathcal{V}$ is the support of the full conditional distribution of $(V_0,V,U,T,R)$. The fact that $c_l>0$, for all $l$, completes the proof.

\section*{Appendix B - MCMC details}

\begin{algorithm}[!h]
\caption{MH step for $(V_0,V,R)$}\label{alg_1}\scriptsize{
\begin{algorithmic}[!h]
\Require{$(V_0,V,R)$, $y$, $E$, $\pi_0$, $g_k$, $\psi_k$'s, $W$, $B$.}
\Ensure{$(V_0,V,R)$.}
\State Compute $c = (c_0, \cdots, c_{|W|})$ from (\ref{eq_propMH}) for the current values of $(V_0,V,R)$.
\For{$j = 1 \to E$}
\State Compute $c_0^*(V_0 = j)$.
\EndFor
\State \textbf{end for}
\State Compute $c_0^*$ and define\\ $p = c_0^* \left(\frac{\pi_0(V_0 = 1)}{c_0(V_0 = 1)}, \cdots, \frac{\pi_0(V_0 = E)}{c_0(V_0 = E)} \right)$.
\State Sample $V_0^* \sim \mbox{Multinomial}(1, p)$.
\State Sample $R_0^*$ from the density or probability vector $c_0(V_0)L_0(V_0, R_0)\pi(R_0|V_0)$.
\For{$l = 1 \to |W|$}
\For{$j = 1 \to E$}
\State Compute $c_l^*(V_l = j)$ (if $V_{l-1}=V_l$, $c_l^*(V_l = j)=1$).
\EndFor
\State \textbf{end for}
\State Compute $c_l^*$ and define\\ $p = c_l^* \left(\frac{\pi(V_l = 1|V_{l-1},\theta)}{c_l(V_l = 1)}, \cdots, \frac{\pi(V_l = E|V_{l-1}, \theta)}{c_l(V_l = E)} \right)$.
\State Sample $V_l^* \sim \mbox{Multinomial}(1, p)$.
\State If $V_{l-1} \neq V_l$, sample $R_(l)^*$ from the density or probability vector $c_l(V_l)L_l(V_{0:l}, R_{0:(l)})\pi(R_{(l)}|V_l)$.
\EndFor
\State \textbf{end for}
\State Sample $u \sim \mbox{Uniform}(0,1)$.
\If{$u < \left( 1 \wedge \prod_{j = 1}^{|W|} \frac{c_j}{c_j^*}\right)$} \Return $(V_0^*,V^*,\cdots,R^*)$.
\Else \ \Return $(V_0,V,R)$.
\EndIf
\end{algorithmic}}
\end{algorithm}

\pagebreak
\section*{Appendix C - Results from the simulations}

\begin{table*}[h!]
  \caption{Posterior statistics of the parameters for scenarios A in Section \ref{subseccompcost}.}\label{tab_ap_1}
  \centering\scriptsize{
\begin{tabular*}{\textwidth}{@{\extracolsep{\fill}}c|cc|ccc}
\hline
scen. & Parameter   & True value & Mean & SD & CI95\% \\ \hline
   & $\psi_{11}$ & 0.5   & 1.38  & 1.02 & (0.08,3.91) \\
   & $\psi_{12}$ & 0.25  & 0.38  & 0.28 & (0.02,1.15) \\
A1 & $\psi_{21}$ & 6     & 5.84  & 0.92 & (4.18,7.81) \\
   & $\psi_{22}$ & -0.25 & -0.27 & 0.11 & (-0.50,-0.06) \\
   & $\psi_{31}$ & 0.5   & 0.60  & 0.25 & (0.22,1.04) \\ \hline
   & $\psi_{11}$ & 2.50  & 3.07  & 2.31 & (0.16,8.79) \\
   & $\psi_{12}$ & 1.25  & 1.45  & 0.29 & (0.94,2.10) \\
A2 & $\psi_{21}$ & 30.00 & 29.86 & 2.63 & (24.83,35.13) \\
   & $\psi_{22}$ & -1.25 & -1.20 & 0.30 & (-1.78,-0.64) \\
   & $\psi_{31}$ & 2.50  & 2.35  & 0.35 & (1.71,3.07) \\ \hline
   & $\psi_{11}$ & 10    & 13.75 & 6.67 & (2.47,27.43) \\
   & $\psi_{12}$ & 5     & 4.50  & 0.52 & (3.41,5.45) \\
A3 & $\psi_{21}$ & 120   & 119.15& 5.37 & (108.83,129.65) \\
   & $\psi_{22}$ & -5    & -4.68 & 0.60 & (-5.85,-3.50) \\
   & $\psi_{31}$ & 10    & 9.03  & 0.64 & (7.82,10.31) \\ \hline
   & $\psi_{11}$ & 50    & 40.81 & 13.65& (19.79,67.81) \\
   & $\psi_{12}$ & 25    & 24.69 & 0.94 & (22.83,26.46) \\
A4 & $\psi_{21}$ & 600   & 597.80& 11.89& (574.64,621.42) \\
   & $\psi_{22}$ & -25   & -24.88& 1.30 & (-27.45,-22.31) \\
   & $\psi_{31}$ & 50    & 47.66 & 1.47 & (44.83,50.59) \\ \hline
   & $\psi_{11}$ & 150   & 202.53& 24.65& (129.16,240.98) \\
   & $\psi_{12}$ & 75    & 72.54 & 1.84 & (68.95,76.17) \\
A5 & $\psi_{21}$ & 1800  &1800.26& 20.91& (1759.32,1840.96) \\
   & $\psi_{22}$ & -75   & -76.16& 2.29 & (-80.55,-71.65) \\
   & $\psi_{31}$ & 150   & 144.11& 2.48 & (139.27,149.01) \\ \hline
\end{tabular*}}
\end{table*}

\begin{figure}[h!]
	\centering{
		\includegraphics[width=1\linewidth]{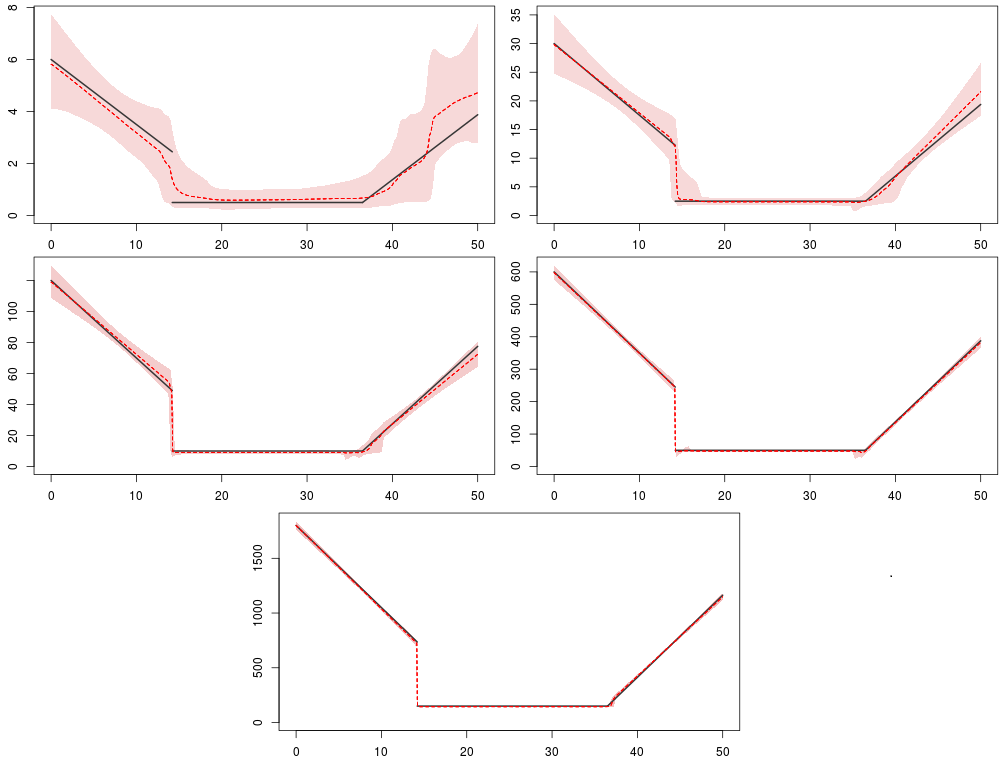}
		\caption{True and estimated (posterior mean and pointwise 95\% CI) intensity function for scenarios A in Section \ref{subseccompcost}, ordered by row.}
		\label{fig_ap_1}}
\end{figure}

\begin{table}[h]
  \caption{Posterior statistics for the number of CTMC jumps ($\#$jumps) and time of the first change in the IF ($T_1$) for scenarios A in Section \ref{subseccompcost}.}\label{tab_ap_4}
  \centering\scriptsize{
\begin{tabular}{cc|c|cc}
\hline
   &           & True value & Mean & SD \\ \hline
A1 & $\#$jumps & 2     & 2.68   & 0.93  \\
   & $T_1$     & 14.2  & 14.28  & 3.74  \\ \hline
A2 & $\#$jumps & 2     & 2.29   & 0.59 \\
   & $T_1$     & 14.2  & 14.43  & 1.1 \\ \hline
A3 & $\#$jumps & 2     & 2.13   & 0.45 \\
   & $T_1$     & 14.2  & 14.13  & 0.74 \\ \hline
A4 & $\#$jumps & 2     & 2.16   & 0.45 \\
   & $T_1$     & 14.2  & 14.18  & 0.43  \\ \hline
A5 & $\#$jumps & 2     & 2.02   & 0.20 \\
   & $T_1$     & 14.2  & 14.18  & 0.42 \\ \hline
\end{tabular}}
\end{table}

\begin{table*}[h]
  \caption{Posterior statistics of the parameters for scenarios B in Section \ref{subseccompcost}.}\label{tab_ap_2}
  \centering\scriptsize{
\begin{tabular*}{\textwidth}{@{\extracolsep{\fill}}c|cc|ccc}
\hline
scen. & Parameter   & True value & Mean & SD & CI95\% \\ \hline
  & $\psi_{11}$ &1.7    & 2.07 & 0.97 & (0.78,4.81) \\
  & $\psi_{12}$ & 0.85  & 0.86 & 0.07 & (0.72,0.98) \\
  & $\psi_{21}$ & 20.4  & 19.47& 1.15 & (17.20,21.68) \\
  & $\psi_{22}$ & -0.85 & -0.79& 0.10 & (-0.97,-0.57) \\
  & $\psi_{31}$ & 1.7   & 1.50 & 0.21 & (1.11,1.95) \\
B1& $\theta_1$  & 0.066 & 0.07 & 0.03 & (0.02,0.14)  \\
  & $\theta_2$  & 0.066 & 0.09 & 0.04 & (0.03,0.17)  \\
  & $\theta_3$  & 0.066 & 0.07 & 0.05 & (0.01,0.20)  \\
  & $(\theta_{11},\theta_{12},\theta_{13})$  & (1/3,1/3,1/3) & (0.24,0.50,0.26) & (0.16,0.18,0.17) & \\
  & $(\theta_{21},\theta_{22},\theta_{23})$  & (1/3,1/3,1/3) & (0.39,0.32,0.29) & (0.22,0.17,0.20) & \\
  & $(\theta_{31},\theta_{32})$            & (0.5,0.5)     & (0.71,0.29)      & (0.22,0.22) & \\ \hline
  & $\psi_{11}$ & 1     & 1.15 & 0.52 & (0.44,2.57) \\
  & $\psi_{12}$ & 0.5   & 0.49 & 0.03 & (0.43,0.55) \\
  & $\psi_{21}$ & 12    &11.68 & 0.67 & (10.36,12.96) \\
  & $\psi_{22}$ & -0.5  & -0.47& 0.06 & (-0.59,-0.35) \\
  & $\psi_{31}$ & 1     & 1.07 & 0.12 & (0.84,1.29) \\
B2& $\theta_1$  & 0.066 & 0.07 & 0.02 & (0.03,0.13)  \\
  & $\theta_2$  & 0.066 & 0.08 & 0.03 & (0.03,0.15)  \\
  & $\theta_3$  & 0.066 & 0.09 & 0.04 & (0.04,0.19)  \\
  & $(\theta_{11},\theta_{12},\theta_{13})$  & (1/3,1/3,1/3) & (0.23,0.22,0.55) & (0.15,0.13,0.17) & \\
  & $(\theta_{21},\theta_{22},\theta_{23})$  & (1/3,1/3,1/3) & (0.37,0.25,0.38) & (0.21,0.16,0.21) & \\
  & $(\theta_{31},\theta_{32})$            & (0.5,0.5)     & (0.72,0.28)      & (0.16,0.16) & \\ \hline
  & $\psi_{11}$ &  0.5  & 0.68 & 0.22 & (0.29,1.15) \\
  & $\psi_{12}$ &  0.25 & 0.25 & 0.02 & (0.20,0.29) \\
  & $\psi_{21}$ & 6     & 5.88 & 0.26 & (5.39,6.39) \\
  & $\psi_{22}$ & -0.25 & -0.24& 0.02 & (-0.28,-0.20) \\
  & $\psi_{31}$ & 0.5   & 0.43 & 0.06 & (0.32,0.54) \\
B3& $\theta_1$  & 0.066 & 0.09 & 0.02 & (0.05,0.15) \\
  & $\theta_2$  & 0.066 & 0.09 & 0.02 & (0.05,0.13) \\
  & $\theta_3$  & 0.066 & 0.07 & 0.03 & (0.03,0.14) \\
  & $(\theta_{11},\theta_{12},\theta_{13})$  & (1/3,1/3,1/3) & (0.12,0.47,0.41) & (0.09,0.14,0.14) & \\
  & $(\theta_{21},\theta_{22},\theta_{23})$  & (1/3,1/3,1/3) & (0.55,0.25,0.20) & (0.17,0.14,0.14) & \\
  & $(\theta_{31},\theta_{32})$            & (0.5,0.5)     & (0.67,0.33)      & (0.18,0.18) & \\ \hline
\end{tabular*}}
\end{table*}

\begin{figure}[h!]
	\centering{
		\includegraphics[width=1\linewidth]{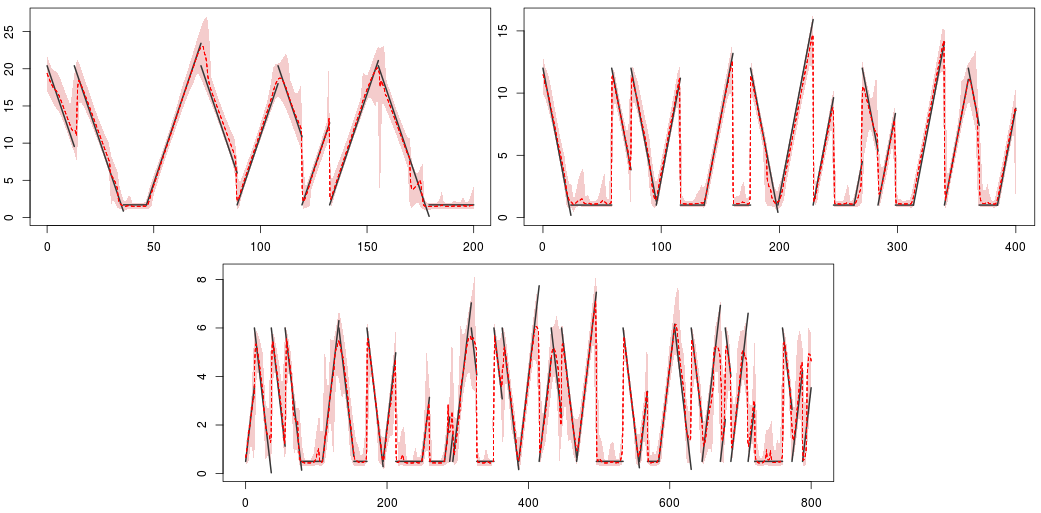}
		\caption{True and estimated (posterior mean and pointwise 95\% CI) intensity function for scenarios B in Section \ref{subseccompcost}, ordered by row.}
		\label{fig_ap_2}}
\end{figure}

\begin{table}[h]
  \caption{Posterior statistics for the number of CTMC jumps ($\#$jumps) and time of the first change in the IF ($T_1$) for scenarios B in Section \ref{subseccompcost}.}\label{tab_ap_5}
  \centering\scriptsize{
\begin{tabular}{cc|c|cc}
\hline
   &           & True value & Mean & SD \\ \hline
B1 & $\#$jumps & 10    & 14.23  & 2.48  \\
   & $T_1$     & 12.8  & 8.23   & 5.47  \\ \hline
B2 & $\#$jumps & 20    & 30.34  & 5.96 \\
   & $T_1$     & 23.6  & 16.53  & 8.94 \\ \hline
B3 & $\#$jumps & 40    & 67.0   & 9.14 \\
   & $T_1$     & 12.4  & 10.03  & 5.56 \\ \hline
\end{tabular}}
\end{table}

\begin{table*}[h]
  \caption{Posterior statistics of the parameters for the examples in Section \ref{subsecpriorsenst}. Inf. and N-inf refer to informative and non-informative priors, respectively. }\label{tab_ap_3}
  \centering\scriptsize{
\begin{tabular*}{\textwidth}{@{\extracolsep{\fill}}c|cc|cc|cc|cc}
\hline
scen. & Param.      & True  & \multicolumn{2}{c|}{Mean} & \multicolumn{2}{c|}{SD} & \multicolumn{2}{c|}{CI95\%} \\ \hline
      &             &       & Inf. & N-inf.& Inf. & N-inf. & Inf. & N-inf.\\ \hline
      & $\psi_{11}$ & 0.5   & 1.38 & 2.35 & 1.02 & 2.81 & (0.08,3.91)   & (0.11,10.52) \\
      & $\psi_{12}$ & 0.25  & 0.38 & 0.47 & 0.28 & 0.52 & (0.02,1.15)   & (0.03,2.20)  \\
 A1   & $\psi_{21}$ & 6     & 5.84 & 5.73 & 0.92 & 0.96 & (4.18,7.81)   & (4.05,7.77)  \\
      & $\psi_{22}$ & -0.25 &-0.27 &-0.26 & 0.11 & 0.12 &-(0.50,-0.06)  & (-0.47,-0.05)\\
      & $\psi_{31}$ & 0.5   & 0.60 & 0.69 & 0.25 & 0.18 & (0.22,1.04)    & (0.36,1.07)  \\ \hline
      & $\psi_{11}$ & 10    &11.03 &13.75 & 4.25 & 6.67 & (3.29,19.02)  & (2.47,27.43) \\
      & $\psi_{12}$ & 5     & 4.68 & 4.50 & 0.42 & 0.52 & (3.88,5.50)   & (3.41,5.45)  \\
 A3   & $\psi_{21}$ & 120   &119.4 &119.15 & 5.18 & 5.37 &(109.3,129.7)  & (108.83,139.65)\\
      & $\psi_{22}$ & -5    &-4.72 &-4.68 & 0.59 & 0.60 &(-5.86,-3.56)  & (-5.85,-3.50)\\
      & $\psi_{31}$ & 10    & 8.99 & 9.03 & 0.63 & 0.64 & (7.79,10.28)  & (7.82,10.31) \\ \hline
      & $\psi_{11}$ & 150   &190.7 &202.53 &28.20 &24.65 &(120.8,230.7)  & (129.16,240.98)\\
      & $\psi_{12}$ & 75    &73.1  & 72.54 & 1.78 & 1.84 &(69.5,76.5)    & (68.95,76.17)  \\
 A5   & $\psi_{21}$ & 1800  &1800.6&1800.26&20.23 &20.91 &(1760.5,1840.3)& (1759.32,1840.96)\\
      & $\psi_{22}$ & -75   &-76.2 &-76.16 & 2.22 & 2.29 &(-80.5,-71.8)  & (-80.55,-71.65)\\
      & $\psi_{31}$ & 150   &144.1 &144.11& 2.47 & 2.48 &(139.3,149.0)  & (139.27,149.01)\\ \hline
      & $\psi_{11}$ & 1     & 1.23 & 1.15 & 0.51 & 0.52 & (0.51,2.54)   & (0.44,2.57)  \\
      & $\psi_{12}$ & 0.5   & 0.49 & 0.49 & 0.03 & 0.03 & (0.43,0.55)   & (0.43,0.55)  \\
 B2   & $\psi_{21}$ & 12    & 11.67&11.68 & 0.66 & 0.67 &(10.37,12.97)  & (10.36,12.96)\\
      & $\psi_{22}$ & -0.5  &-0.47 &-0.47 & 0.06 & 0.06 &(-0.60,-0.35)  & (-0.59,-0.35)\\
      & $\psi_{31}$ & 1     & 1.07 & 1.07 & 0.12 & 0.12 & (0.84,1.29)   & (0.84,1.29)  \\
      & $\theta_1$  & 0.066 & 0.07 & 0.07 & 0.02 & 0.02 & (0.03,0.12)   & (0.03,0.12)  \\
      & $\theta_2$  & 0.066 & 0.08 & 0.08 & 0.03 & 0.03 & (0.03,0.15)   & (0.03,0.15)  \\
      & $\theta_3$  & 0.066 & 0.09 & 0.09 & 0.04 & 0.04 & (0.04,0.19)   & (0.04,0.19)  \\ \hline
\end{tabular*}}
\end{table*}

\begin{figure}[h!]
	\centering{
		\includegraphics[width=1\linewidth]{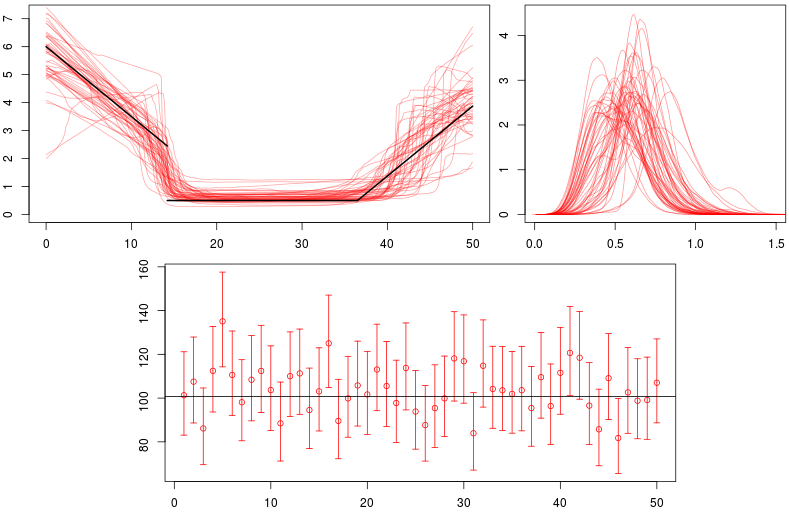}
		\caption{Results for 50 replications of scenario A1. Top left: real and posterior mean of the IF for each replication. Top right: Posterior density of the measure of fit in (\ref{meas_fit}) for each replication. Bottom: mean and 95\% CI for the integrated IF for each replication. Horizontal line represents the real value.}
		\label{fig_ap_4}}
\end{figure}
\begin{figure}[h!]
	\centering{
		\includegraphics[width=1\linewidth]{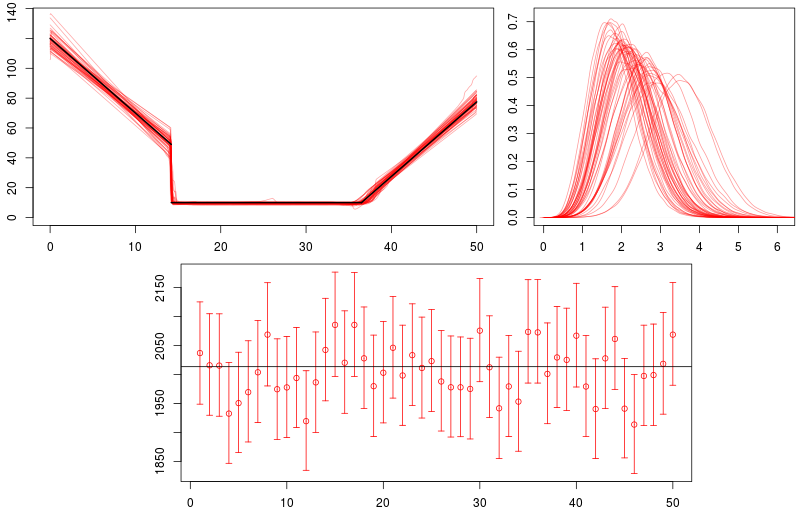}
		\caption{Results for 50 replications of scenario A3. Top left: real and posterior mean of the IF for each replication. Top right: Posterior density of the measure of fit in (\ref{meas_fit}) for each replication. Bottom: mean and 95\% CI for the integrated IF for each replication. Horizontal line represents the real value.}
		\label{fig_ap_5}}
\end{figure}
\begin{figure}[h!]
	\centering{
		\includegraphics[width=1\linewidth]{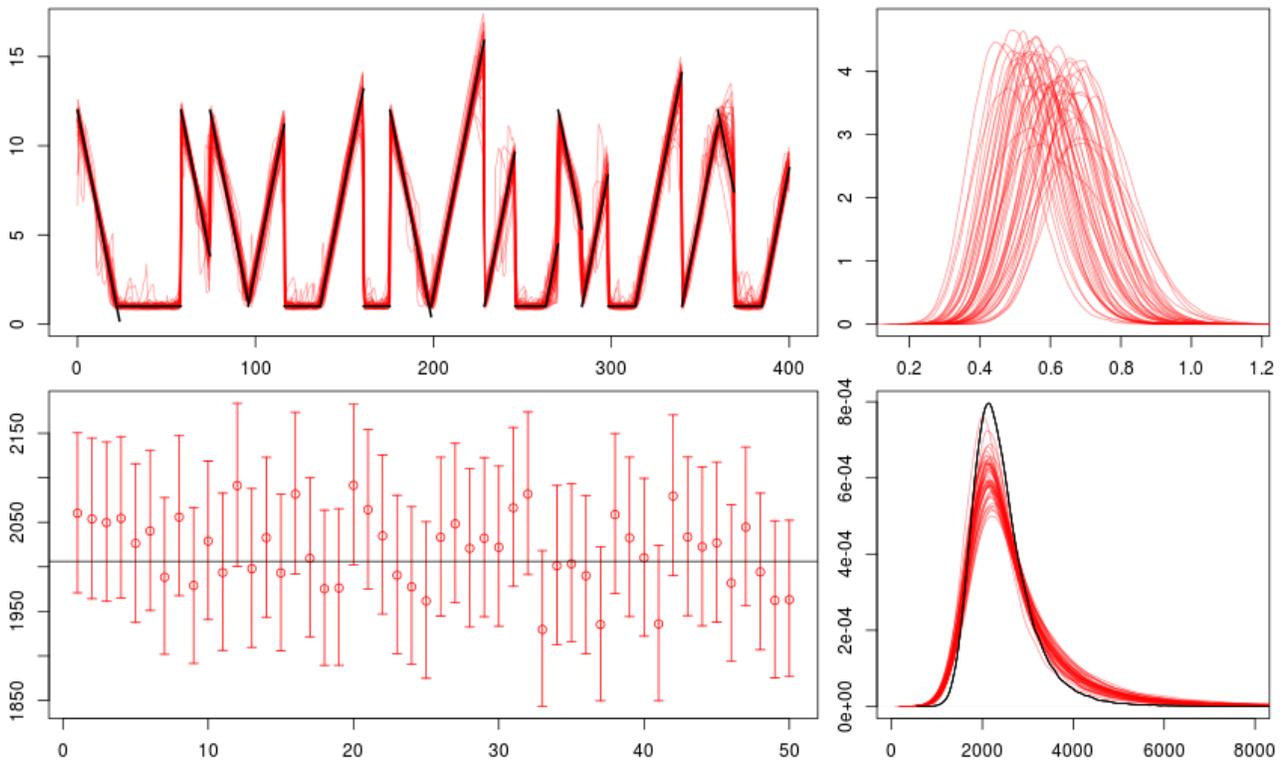}
		\caption{Results for 50 replications of scenario B2. Top left: real and posterior mean of the IF for each replication. Top right: Posterior density of the measure of fit in (\ref{meas_fit}) for each replication. Bottom: mean and 95\% CI for the integrated IF for each replication. Horizontal line represents the real value. Bottom right: real (black line) and posterior predictive density of the integrated IF in $[400,800]$.}
		\label{fig_ap_6}}
\end{figure}
\begin{figure}[h!]
	\centering{
		\includegraphics[width=0.8\linewidth]{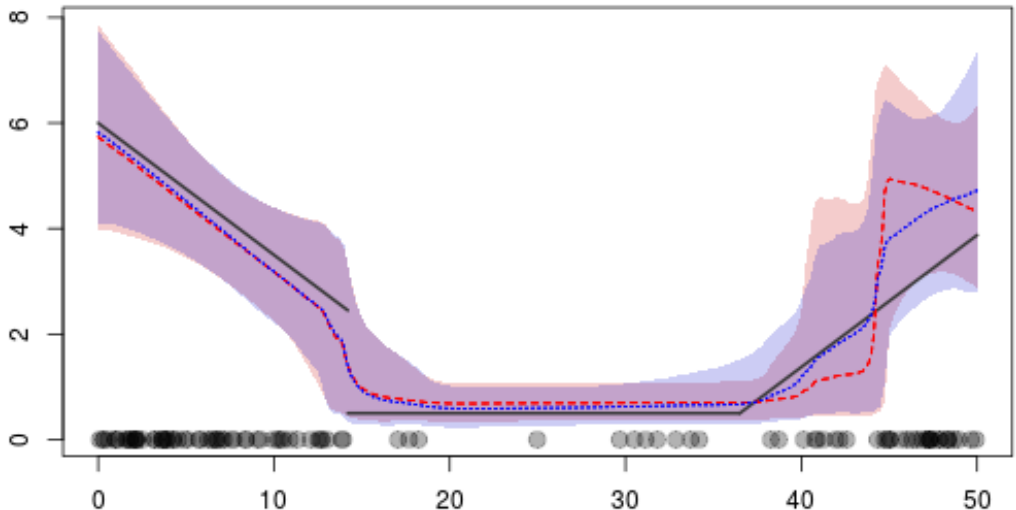}
		\caption{True and estimated - posterior mean and pointwise 95\% CI, intensity function for scenarios A1 in Section \ref{subsecpriorsenst} for the informative (blue) and non-informative (red) priors on $\psi$.}
		\label{fig_ap_3}}
\end{figure}
\begin{figure}[h!]
	\centering{
		\includegraphics[width=0.8\linewidth]{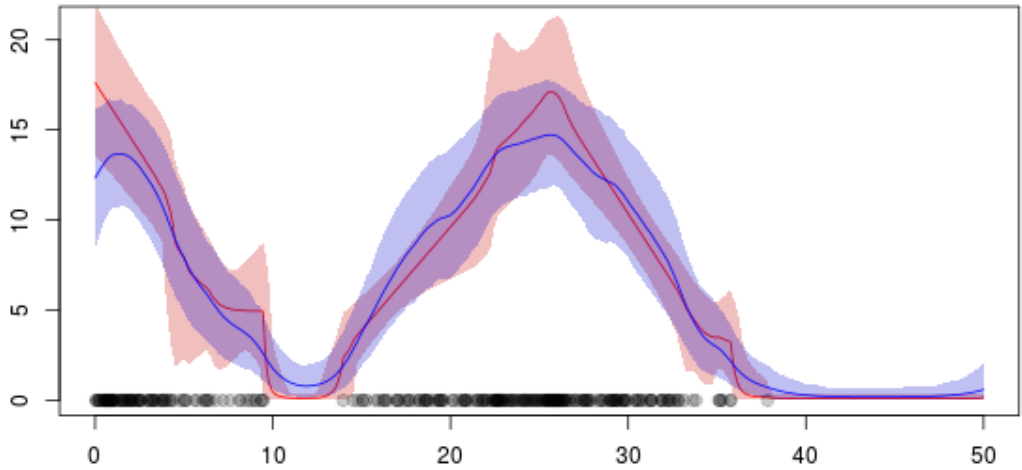}
		\caption{Comparison between the estimated IF (posterior mean and 95\% pointwise CI) for the GMMPP (red) and the non-parametric IF (blue) models.}
		\label{fig_ap_7}}
\end{figure}
\begin{figure}[h!]
	\centering{
		\includegraphics[width=0.5\linewidth]{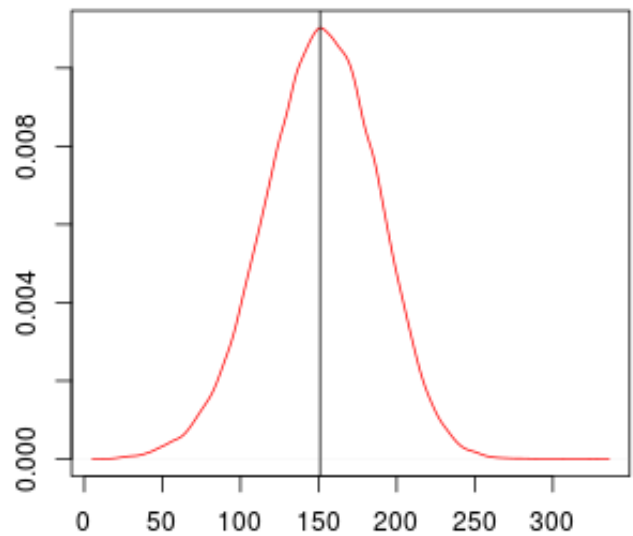}
		\caption{Posterior predictive density of the integrated IF for the exchange rate example. The vertical line represents the real observed number of events in the predicted period.}\label{fig_ap_8}}
\end{figure}

\newpage

\textcolor[rgb]{1.00,1.00,1.00}{.}

\newpage

\textcolor[rgb]{1.00,1.00,1.00}{.}

\newpage

\textcolor[rgb]{1.00,1.00,1.00}{.}

\newpage

\textcolor[rgb]{1.00,1.00,1.00}{.}

\newpage

\textcolor[rgb]{1.00,1.00,1.00}{.}

\newpage

\section*{Appendix D - MCMC diagnostics}

\begin{figure}[h!]
	\centering{
		\includegraphics[width=0.5\linewidth]{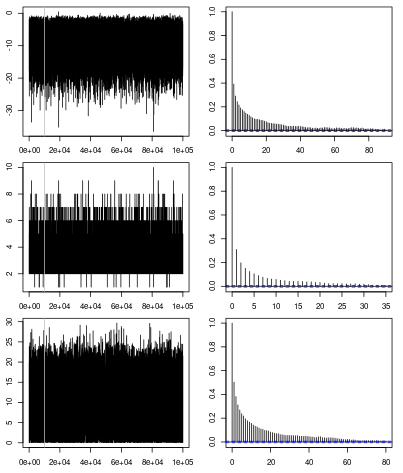}
		\caption{Trace plot and autocorrelation plot for the log-posterior density, number of CTMC jumps and time of the first change in the IF, respectively, for scenario A1 in Section \ref{subseccompcost}. }
		\label{fig_ap_9}}
\end{figure}
\begin{figure}[h!]
	\centering{
		\includegraphics[width=0.5\linewidth]{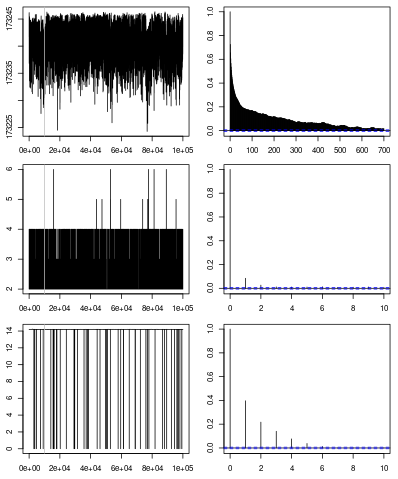}
		\caption{Trace plot and autocorrelation plot for the log-posterior density, number of CTMC jumps and time of the first change in the IF, respectively, for scenario A5 in Section \ref{subseccompcost}.}
		\label{fig_ap_13}}
\end{figure}
\begin{figure}[h!]
	\centering{
		\includegraphics[width=0.5\linewidth]{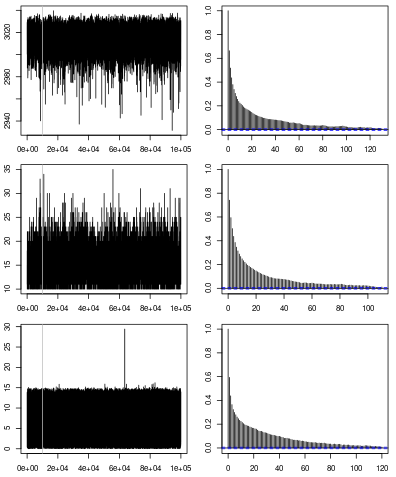}
		\caption{Trace plot and autocorrelation plot for the log-posterior density, number of CTMC jumps and time of the first change in the IF, respectively, for scenario B1 in Section \ref{subseccompcost}.}
		\label{fig_ap_14}}
\end{figure}
\begin{figure}[h!]
	\centering{
		\includegraphics[width=0.5\linewidth]{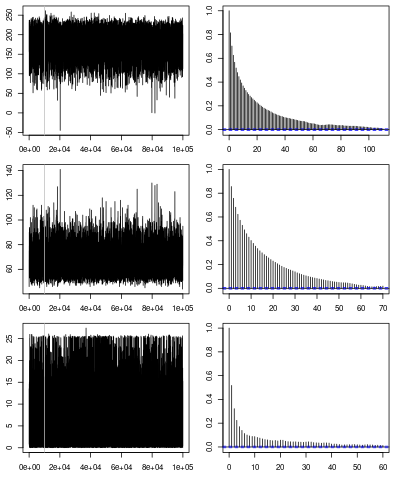}
		\caption{Trace plot and autocorrelation plot for the log-posterior density, number of CTMC jumps and time of the first change in the IF, respectively, for scenario B3 in Section \ref{subseccompcost}.}
		\label{fig_ap_16}}
\end{figure}

\end{document}